\documentclass[iop,apj,tighten]{emulateapj}
\usepackage{apjfonts} 
\usepackage{amsmath,amstext}
\usepackage[breaklinks,colorlinks,citecolor=blue,linkcolor=magenta]{hyperref} 
\usepackage[all]{hypcap} 

\shorttitle{Jet formation}
\shortauthors{Gonz\'alez-Avil\'es et al.}

\begin{document}

\title{Jet formation in solar atmosphere due to magnetic reconnection}

\author{J.J. Gonz\'alez-Avil\'es \altaffilmark{1} , F.S. Guzm\'an  \altaffilmark{1}, and V. Fedun \altaffilmark{2}}
\affil{$^1$Laboratorio de Inteligencia Artificial y Superc\'omputo. Instituto de F\'{\i}sica y Matem\'{a}ticas, Universidad
        Michoacana de San Nicol\'as de Hidalgo. Morelia, Michoac\'{a}n, M\'{e}xico.\\
        $^2$ Space Systems Laboratory, Department of Automatic Control and Systems Engineering, University of Sheffield, S1 3JD, UK}

\begin{abstract}
Using numerical simulations, we show that jets with features of type II spicules and cold coronal jets corresponding to temperatures $10^{4}$ K can be formed due to magnetic reconnection in a scenario in presence of magnetic resistivity. For this we model the low chromosphere-corona region using the C7 equilibrium solar atmosphere model and assuming Resistive MHD rules the dynamics of the plasma. The magnetic filed configurations we analyze correspond to two neighboring loops with opposite polarity. The separation of the loops' feet determines the thickness of a current sheet that triggers a magnetic reconnection process, and the further formation of a high speed and sharp structure. We analyze the cases where the magnetic filed strength of the two loops is equal and different. In the first case, with a symmetric configuration the spicules raise vertically whereas in an asymmetric configuration the structure shows an inclination. With a number of simulations carried out under a 2.5D approach, we explore various properties of excited jets, namely, the morphology,  inclination and velocity. The parameter space involves magnetic field strength between 20 and 40 G, and the resistivity is assumed to be uniform with a constant value of the order $10^{-2}\Omega\cdot m$.
\end{abstract}



\section{Introduction}
\label{sec:Introduction}

Magnetic reconnection is a topological reconfiguration of the magnetic field caused by changes in the connectivity of its field lines \citep{Priest_1984,Priest_et_al_2000}. It is also a mechanism of conversion of magnetic energy into thermal and kinetic energy of plasma when two antiparallel magnetic fields encounter and reconnect with each other. Magnetic reconnection can occur in the chromosphere, photosphere and even in the convection zone. In particular, the chromosphere has a very dynamical environment  where magnetic features such as H$_\alpha$ upward flow events \citep{Chae_et_al_1998} and erupting mini-filaments \citep{Wang_et_al_2000} happen. The dynamics of the chromosphere at the limb region is dominated by spicules \citep{Beckers_1968} and related flows such as mottles and fibrils on the disk \citep{Hansteen_et_al_2006,De_Pontieu_et_al_2007a}. Spicular structures are also visible at the limb in many spectral lines at the transition region temperatures \citep{Mariska_1992,Wilhelm_2000}, and some observations suggest that coronal dynamics are linked to spicule-like jets \citep{McIntosh_et_al_2007,Tsiropoula_et_al_2012,Cheung_et_al_2015,Skogsrud_et_al_2015,Tavabi_et_al_2015a,Narang_et_al_2016}.  With the large improvement in spatiotemporal stability and resolution given by the Hinode satellite \citep{Kosugi_et_al_2007}, and with the Swedish 1 m Solar Telescope (SST) \citep{Scharmer_et_al_2008}, two classes of spicules were defined in terms of their different dynamics and timescales \citep{De_Pontieu_et_al_2007c}.

The so-called type I spicules have lifetimes of 3 to 10 minutes, achieve speeds of 10-30 km s$^{-1}$, and reach heights of 2-9 Mm \citep{Beckers_1968,Suematsu_et_al_1995}, and typically involve upward motion followed by downward motion. In \citep{Shibata&Suematsu_1982,Shibata_et_al_1982}, the authors studied in detail the propagating shocks using simplified one-dimensional models. In \citep{Hansteen_et_al_2006,De_Pontieu_et_al_2007a,Heggland_et_al_2007}, the authors study the propagation of shocks moving upwards passing through the upper chromosphere and transition region toward the corona. They also describe how the spicule-driving shocks can be generated by a variety of processes, such as collapsing granules, p-modes and dissipation of magnetic energy in the photosphere and lower chromosphere. In \citep{Matsumoto&Shibata_2010} the authors state that spicules can be driven by resonant Alfv\'en waves generated in the photosphere and confined in a cavity between the photosphere and the transition region. Another studies in this direction, like \citep{Murawski&Zaqarashvili_2010,Murawski_et_al_2011}, where they use ideal MHD and perturb the velocity field in order to stimulate the formation of type I spicules and macro-spicules. Furthermore in \citep{Scullion_et_al_2011}, the authors simulate the formation of wave-driven type I spicules phenomena in 3D trough a Transition Region Quake (TRQ) and the transmission of acoustic waves from the lower chromosphere to the corona. 

Type II spicules are observed in Ca II and H$\alpha$, these spicules have lifetimes typically less than 100s in contrast with type I spicules that have lifetimes of 3 to 10 min, are more violent, with upward velocities of order 50-100 km s$^{-1}$ and reach greater heights. They usually exhibit only upward motion \citep{De_Pontieu_et_al_2007b}, followed by a fast fading in chromospheric lines without observed downfall. Spicules of  type II seen in the Ca II band of Hinode fade within timescales of the order of a few tenths of seconds \citep{De_Pontieu_et_al_2007a}. The type II spicules observed on the solar disk are dubbed ``Rapid Blueshifted Events'' (RBEs) \citep{Langangen_et_al_2008,Rouppe_et_al_2009}. These show strong Doppler blue shifted lines in  the region from the middle to the upper chromosphere.  The RBEs are linked with asymmetries in the transition region and coronal spectral line profiles \citep{De_Pontieu_et_al_2009}. In addition the lifetime of RBEs suggests that they are heated with at least transition region temperatures \citep{De_Pontieu_et_al_2007c,Rouppe_et_al_2009}. Type II spicules also show transverse motions with amplitudes of 10-30 km s$^{-1}$ and periods of 100-500 s \citep{Tomczyk_et_al_2007,McIntosh_et_al_2011,Zaqarashvili&Erdelyi_2009}, which are interpreted as upward or downward propagating Alfvenic waves \citep{Okamoto&De_Pontieu_2011,Tavabi_et_al_2015b}, or MHD kink mode waves \citep{He_et_al_2009,McLaughlin_et_al_2012,Kuridze_et_al_2012}.  

As mentioned above, there are several theoretical and observational results about type II spicules, but there is little consensus about the origin of type II spicules and the source of their transverse oscillations. Some possibilities discussed suggest that type II spicules are due to the magnetic reconnection process \citep{Isobe_et_al_2008,De_Pontieu_et_al_2007c,Archontis_et_al_2010}, oscillatory reconnection process \citep{McLaughlin_et_al_2012}, strong Lorentz force \citep{Martinez-Sykora_et_al_2011} or propagation of p-modes \citep{de_Wijn_et_al_2009}. Moreover, type II spicules could be warps in 2D sheet like structures \citep{Judge_et_al_2011}.  A more recent study suggests another mechanism, for instance in \citep{Sterling&Moore_2016}, the authors suggested that solar spicules result from the eruptions of small-scale chromospheric filaments.

The limited resolution in observations and the complexity of the chromosphere make difficult the interpretation of the structures, and even question the existence of type II spicules as a particular class, for instance in \citep{Zhang_et_al_2012}. In consequence, these difficulties spoil the potential importance of magnetic reconnection as a transcendent mechanism in the solar surface. Nevertheless, there is evidence that magnetic reconnection is a good explanation of chromospheric anemone jets \citep{Singh_et_al_2012}, which are observed to be much smaller and much more frequent than surges \citep{Shibata_et_al_2007}.  A statistical study performed by \citep{Nishizuka_et_al_2011} showed that the chromospheric anemone jets have typical lengths of 1.0-4.0 Mm, widths of 100-400 km, and cusp size of 700-2000 km. Their lifetimes is about 100-500 s and their velocity is about 5-20 km s$^{-1}$. Other types of coronal jets can be generated by magnetic reconnection, for example in \citep{Yokoyama&Shibata_1995,Yokoyama&Shibata_1996} the authors using two-dimensional numerical simulations study the jet formation, or in \citep{Nishizuka_et_al_2008}, it is shown that emerging magnetic flux reconnects with an open ambient magnetic field and such reconnection produces the acceleration of material and thus a jet structure. The reconnection seems to trigger the jet formation in a horizontally magnetized atmosphere, with the flux emergence as a mechanism \citep{Archontis_et_al_2005,Galsgaard_et_al_2007}. 

Another approach uses a process that produces a magnetic reconnection using numerical dissipation of the ideal MHD equations, and the atmosphere model is limited to have a constant density and pressure profiles and assumes there is no gravity \citep{Pariat_et_al_2009,Pariat_et_al_2010,Pariat_et_al_2015,Rachmeler_et_al_2010}. In this paper we show that magnetic reconnection can be responsible for the formation of  jets with some characteristics of Type II spicules and cold coronal jets \citep{Nishizuka_et_al_2008}, for that i) we solve the system of equations of the Resistive MHD subject the solar gravitational field, ii) we assume a completely ionized solar atmosphere consistent with the C7 model. The resulting magnetic reconnection accelerates the plasma upwards by itself and produces the jet. 

The paper is organized as follows, in Section \ref{sec:model_numerical_methods} we describe the resistive MHD equations, the model of solar atmosphere, the magnetic field configuration used the numerical simulations and the numerical methods we use. In Section \ref{sec:Results}, we present the results of the numerical  simulations for various experiments. Section \ref{sec:conclusions} contains final comments and conclusions. 


\section{Model and Numerical Methods}
\label{sec:model_numerical_methods}


\subsection{The system of Resistive MHD equations}
\label{sub_sec:eglm_resistive_mhd_equations}

The minimum system of equations allowing the formation of magnetic reconnection is the resistive MHD.  In this paper we follow \citep{Jiang_et_al_2012} to write the dimensionless Extended Generalized Lagrange Multiplier (EGLM) resistive MHD equations that include gravity as follows:

\begin{eqnarray}
\frac{\partial\rho}{\partial t} +\nabla\cdot(\rho{\bf{v}})=0, \label{eglm_cont_equation} \\
\frac{\partial(\rho{\bf v})}{\partial t} + \nabla\cdot\left(\left(p+\frac{1}{2}{\bf B}^{2}\right){\bf I}+\rho{\bf vv}-{\bf BB}\right)\\=-(\nabla\cdot{\bf B}){\bf B}+\rho{\bf g}, \label{eglm_mom_equation} \\
\frac{\partial E}{\partial t}+\nabla\cdot\left({\bf v}\left(E+\frac{1}{2}{\bf B}^{2}+p \right)-{\bf B}({\bf B}\cdot{\bf v})\right)\\=-{\bf B}\cdot(\nabla\psi)-\nabla\cdot((\eta\cdot{\bf J})\times{\bf B})+\rho{\bf g}\cdot{\bf v}, \label{eglm_energy_equation} \\[0.3 cm]
\frac{\partial{\bf B}}{\partial t}+\nabla\cdot({\bf Bv}-{\bf vB}+\psi{\bf I})=-\nabla\times(\eta\cdot{\bf J}), \label{eglm_induction_equation} \\
\frac{\partial\psi}{\partial t}+c_{h}^{2}\nabla\cdot{\bf B}=-\frac{c_{h}^{2}}{c_{p}^{2}}\psi, \label{eglm_psi_equation} \\ \nonumber
{\bf J}=\nabla\times{\bf B}, \\ \nonumber
E=\frac{p}{(\gamma-1)}+\frac{\rho{\bf v}^{2}}{2}+\frac{{\bf B}^{2}}{2}, \\ \nonumber
\end{eqnarray}

\noindent where $\rho$ is the mass density, ${\bf v}$ is the velocity vector field, ${\bf B}$ is the magnetic vector field, $E$ is the total energy density, where $\gamma=5/3$, the plasma pressure $p$ is described by the equation of state of an ideal gas; ${\bf g}$ is the gravitational field, ${\bf J}$ is the current density, $\eta$ is the magnetic resistivity tensor and $\psi$ is a scalar potential that helps damping out the violation of the constraint. Here $c_h$ is the wave speed and $c_p$ is the damping rate of the wave of the characteristic mode associated to $\psi$. In this study we consider an uniform and constant magnetic resistivity, because the solar chromosphere is fully collisional and anomalous or space dependent resistivity -which is the result of various collisionless processes- may not be expected \citep{Singh_et_al_2011}. We normalize the equations with the quantities given in Table \ref{table:1}, which are typical scales in the solar corona.   

In the EGLM-MHD formulation, equation (\ref{eglm_psi_equation}) is the magnetic field divergence free constraint. As suggested in \citep{Dedner_2002}, the expressions for $c_h$ and $c_p$ are
 
\begin{equation} 
c_h = \frac{c_{cfl}}{\Delta t}min(\Delta x,\Delta y,\Delta z), 
~~~c_p = \sqrt{-\Delta t\frac{c_{h}^{2}}{\ln c_{d}}}, \nonumber 
\end{equation}

\noindent where $\Delta t$ is the time step, $\Delta x$, $\Delta y$ and $\Delta z$ are the spatial resolutions, $c_{cfl}<1$ is the Courant factor, $c_d$ is a problem dependent coefficient between 0 and 1, this constant determines the damping rate of divergence errors. The parameters $c_h$ and $c_p$ are not independent of the grid resolution and  the numerical scheme used, for that reason one should adjust their values. In our simulations we use $c_p=\sqrt{c_r}c_h$, with $c_r=0.18$ and $c_h=0.01$. For our analysis we use a 2.5D model, but we solve the 3D resistive MHD equations with high resolution in the $xz$ plane and with a 4 cells along the $y$ direction, so that the speeds $c_h$ and $c_p$ only  depend on $\Delta x$ and $\Delta z$. All the state variables depend on $x$, $y$ and $z$ \citep{Gonzalez_A&Guzman_2015,NewtonianCAFE}. 

\begin{table}
\caption{Scaling factors to translate physical to code units}
\centering
\begin{tabular}{c c c c}
\hline\hline
Variable & Quantity & Unit & Value \\ [0.5ex]
\hline
$x$, $y$, $z$ & Length & $l_0$ & $10^{6}$ m \\
$\rho$ & Density & $\rho_{0}$ & $10^{-12}$ kg$\cdot m^{-3}$ \\
${\bf B}$& Magnetic field & $B_{0}$ & 11.21 G \\
${\bf v}$ & Velocity & $v_{0}=B_0/\sqrt{\mu_0\rho_0}$ & $10^{6}$ m$\cdot s^{-1}$ \\
$t$& Time & $t_{0}=l_0/v_0$ & 1 s \\
$\eta$ & Resistivity & $\eta_{0}=l_0\mu_0 v_0$ & 1.25664$\times10^{6} m^{2}\cdot s^{-1}\cdot N\cdot A^{-2}$ \\ [1ex]
\hline
\end{tabular}
\label{table:1}
\end{table}


\subsection{Numerical methods}
\label{sub_sec:numerical_methods}

We solve numerically the resistive EGLM-MHD equations given by the system of equations (\ref{eglm_cont_equation})-(\ref{eglm_psi_equation}) on a single uniform cell centered grid,  using the method of lines with a third order total variation diminishing Runge-Kutta time integrator \citep{Shu&Osher_1989}. In order to use the method of lines, the RHS of resistive MHD equations are discretized using a finite volume approximation with High Resolution Shock Capturing methods \citep{LeVeque_1992}. For this, we first reconstruct the variables at cell interfaces using the Minmod limiter. The numerical fluxes are calculated using the Harten-Lax-van-Leer-Contact (HLLC) approximate Riemann solver \citep{Li_2005}.


\subsection{Model of the solar atmosphere}
\label{sub_sec:solar_atmosphere}

\begin{figure}
\centering
\includegraphics[width=6.5cm,height=5.5cm]{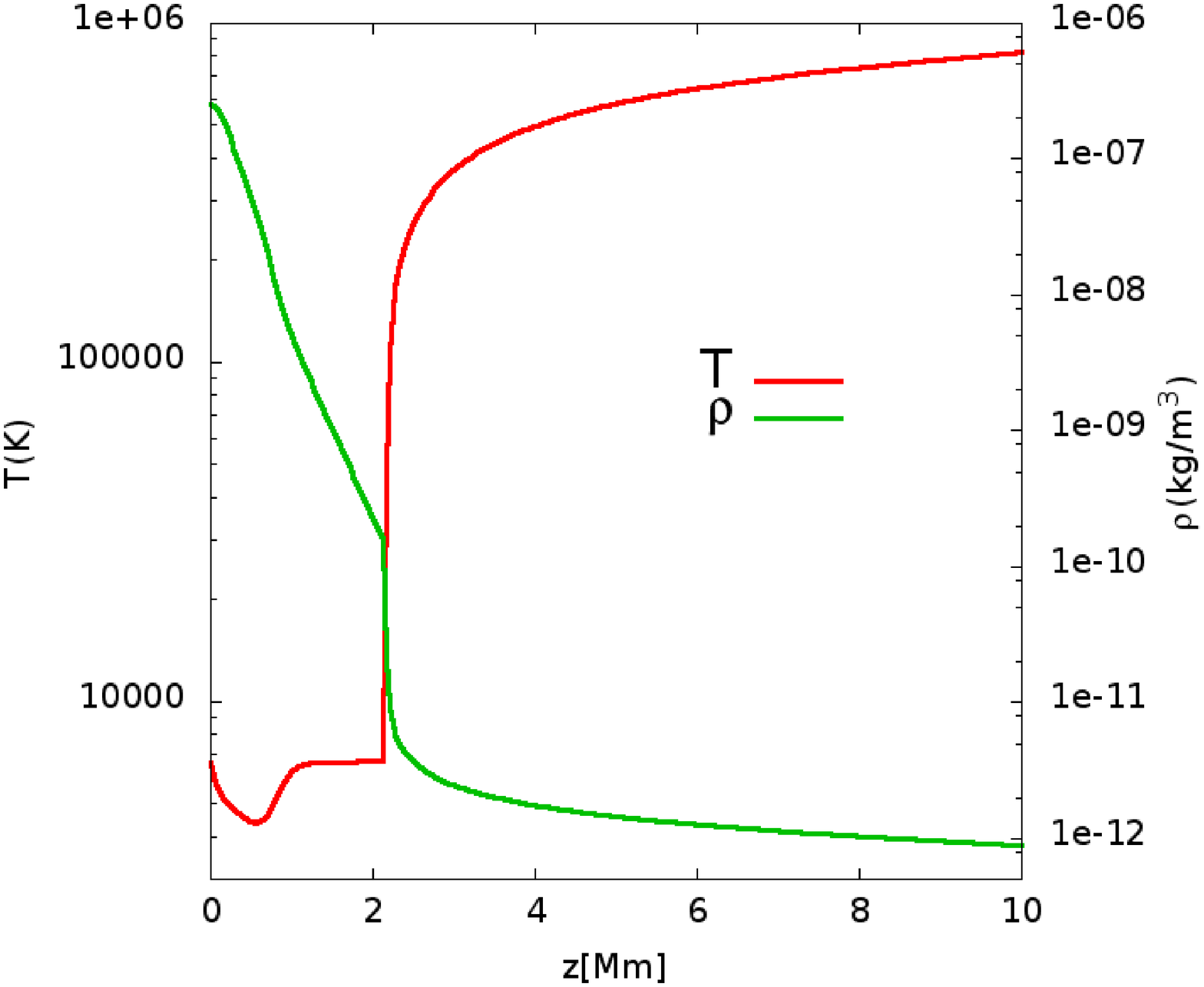}
\caption{Temperature in red and mass density in green as a function of height for the C7 equilibrium solar atmosphere model. Notice the steep jump in Temperature, which makes the C7 model a realistic one.}
\label{fig:atmosphere}
\end{figure}

We choose the numerical domain to cover part of the photosphere, chromosphere and corona. We consider the atmosphere in hydrostatic equilibrium and study the evolution on a finite $xz$ domain, where $x$ is a horizontal coordinate and $z$ labels height. The temperature field is assumed to obey the semiempirical C7 model of the chromosphere \citep{Avrett&Loeser2008} and is distributed to obtain optimum agreement between calculated and observed continuum intensities, line intensities, and line profiles of the SUMER \citep{Curdt_et_al_1999} atlas of the extreme ultraviolet spectrum. The profiles of $T(z)$ and $\rho(z)$ are shown in Fig. \ref{fig:atmosphere}, where the expected gradients at the transition region can be seen.

\subsection{The magnetic field}
\label{sub_sec:initial_conditions}

The magnetic field in the model is chosen as a superposition of two neighboring loops. 
Following \citep{Priest_1982,Del_Zanna_et_al_2005} we construct a loop with the vector potential

\begin{equation}
A_y(x,z) = \frac{B_{01}}{k}\cos(kx)\exp(-kz), \label{vector_potential}
\end{equation} 
 
 \noindent where $B_{01}$ is the photospheric field magnitude at the foot points  $x=\pm L/2$ and $k=\pi/L$. Here $L$ is the distance between the two foot points of the loop and $k$ defines the nodes of the potential. In this model the components of the magnetic field can be represented as
 
  \begin{eqnarray}
B_x(x,z) &=& B_{01}\cos(kx)\exp(-kz), \\
B_z(x,z) &=& -B_{01}\sin(kx)\exp(-kz). 
\end{eqnarray}

\begin{figure}
\centering
\includegraphics[width=8.3cm,height=3.0cm]{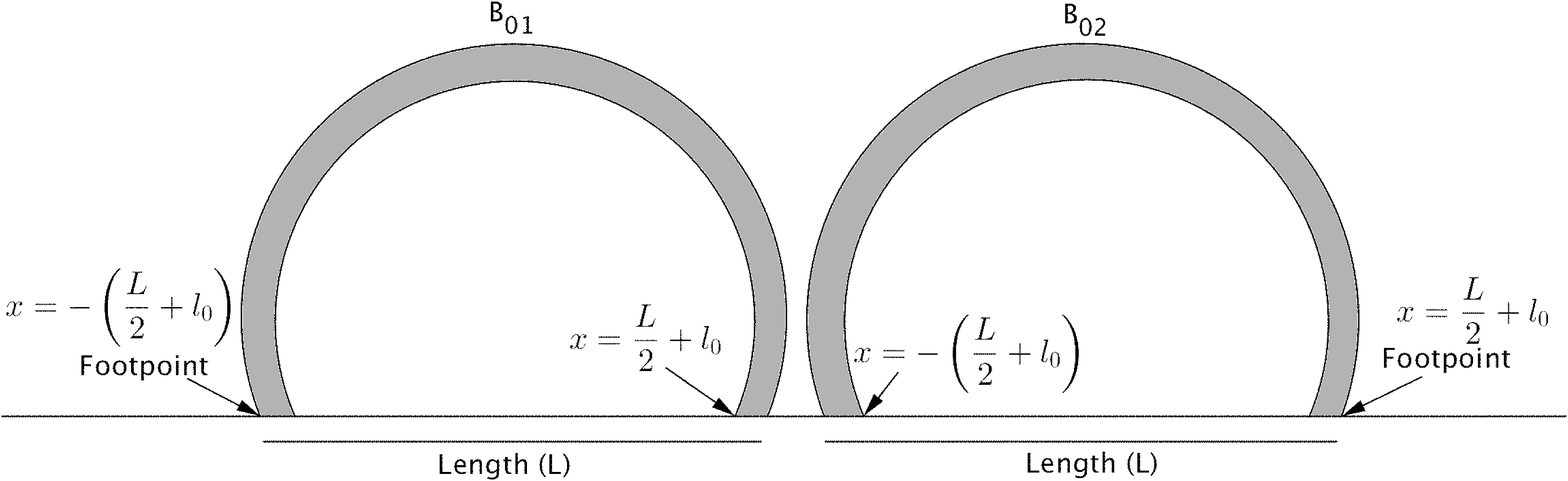}
\includegraphics[width=8.3cm,height=3.0cm]{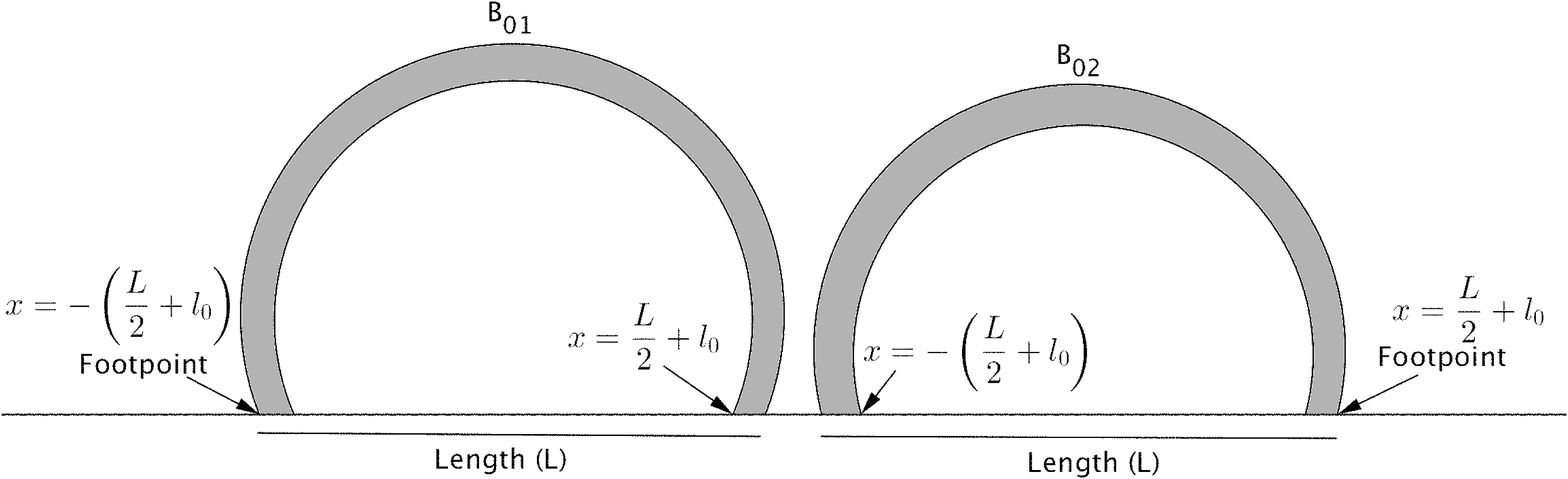}
\caption{\label{fig:magnetic_loops} At the top we show two consecutive symmetric magnetic loop configurations with the same field strength $B_{01}=B_{02}$. At the bottom we show two consecutive non-symmetric magnetic loop configurations for the case $B_{01}>B_{02}$. The length of each loop is represented by $L$ and the location of foot points is determined by the parameter $l_0$.}
\end{figure}

In order to superpose two loops  we use a modified version of (\ref{vector_potential}):

\begin{eqnarray}
A_y(x,z) &=& \frac{B_{01}}{k}\cos(k(x + l_0))\exp(-kz)\nonumber\\ 
&+& \frac{B_{02}}{k}\cos(k(x - l_0))\exp(-kz) , \label{vector_potential_translation}
\end{eqnarray}

\noindent where $l_0$ defines the location of the foot points for each loop, $B_{01}$ and $B_{02}$ are the magnetic field strengths of the left and right loop respectively. With the parameter $l_0$ it is possible to control the separation between the two magnetic loops, which in turn will influence the thickness of a current sheet. The case $B_{01}=B_{02}$ describes two neighboring loop configurations with the same magnetic field strength, whereas  $B_{01} \ne B_{02}$ describes two nearby loops with different magnetic field strength.  A schematic picture of these two configuration is shown in Fig. \ref{fig:magnetic_loops}. 

\section{Results of Numerical Simulations}
\label{sec:Results}

Putting altogether, we run a number of simulations with the magnetic field configuration given by equation (\ref{vector_potential_translation}) in a scenario with constant resistivity $\eta=5\times10^{-2}$ $\Omega\cdot m$ across the whole domain, which is a realistic value estimated for a fully ionized solar atmosphere \citep{Priest_2014}. We experiment with various values of the magnetic field strength and separation of the loops as indicated in Table \ref{tab:CaseB}.  

We fix $k=\pi/L$ with $L=8$ Mm and the simulations were carried out in a domain $x\in[-4,4]$, $y\in[0,1]$, $z\in[0,10]$ in units of Mm, covered with 300$\times$4$\times$375 grid cells. In all the numerical simulations we use out flux boundary conditions, which in our approach translate into copy boundary conditions applied to the conservative variables \citep{Toro}.


\subsection{Symmetric configurations}
\label{sub_sec:symmetric_loop_results}

\begin{figure*}
\centering
\includegraphics[width=4.25cm,height=6.5cm]{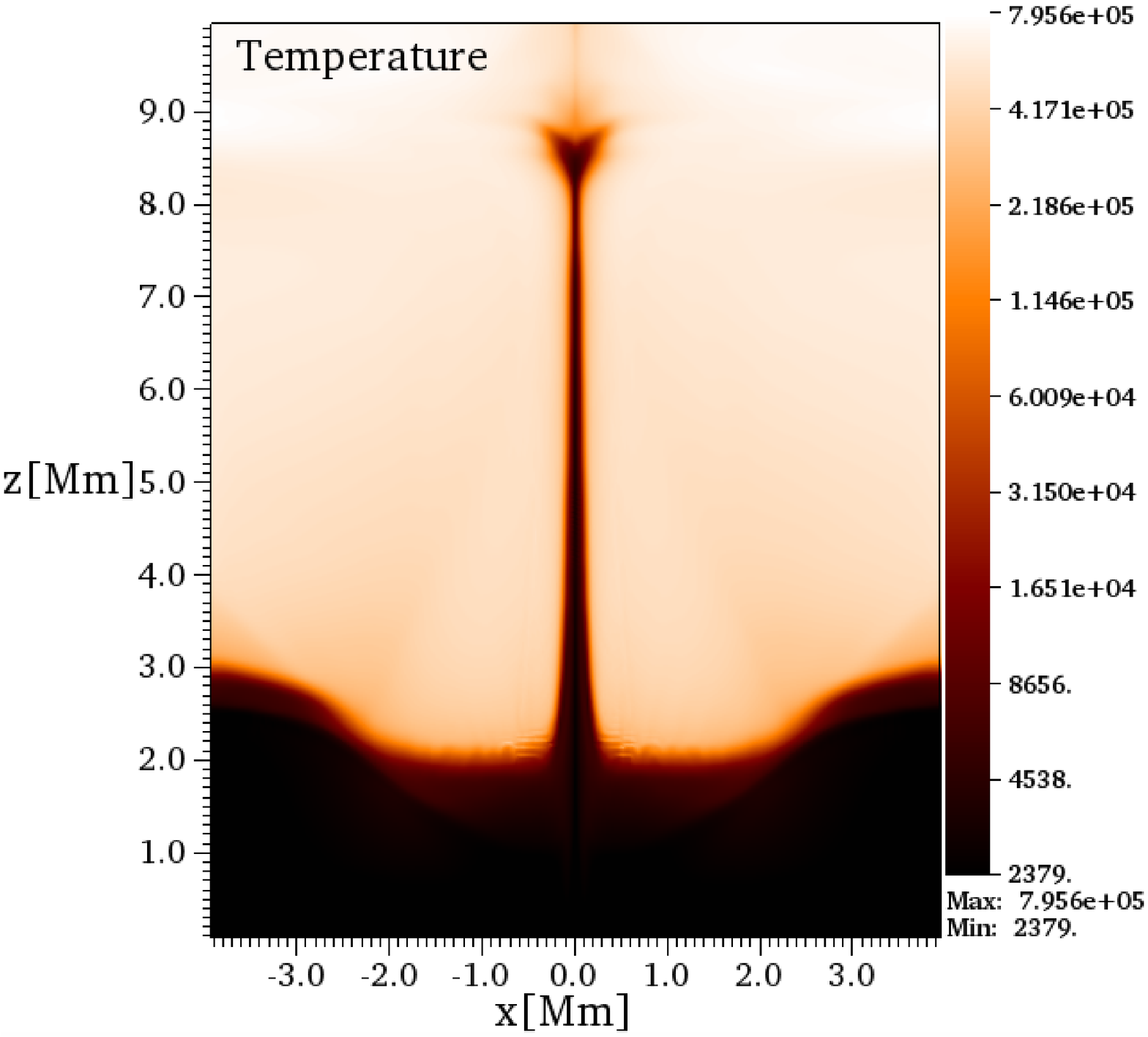}
\includegraphics[width=4.25cm,height=6.5cm]{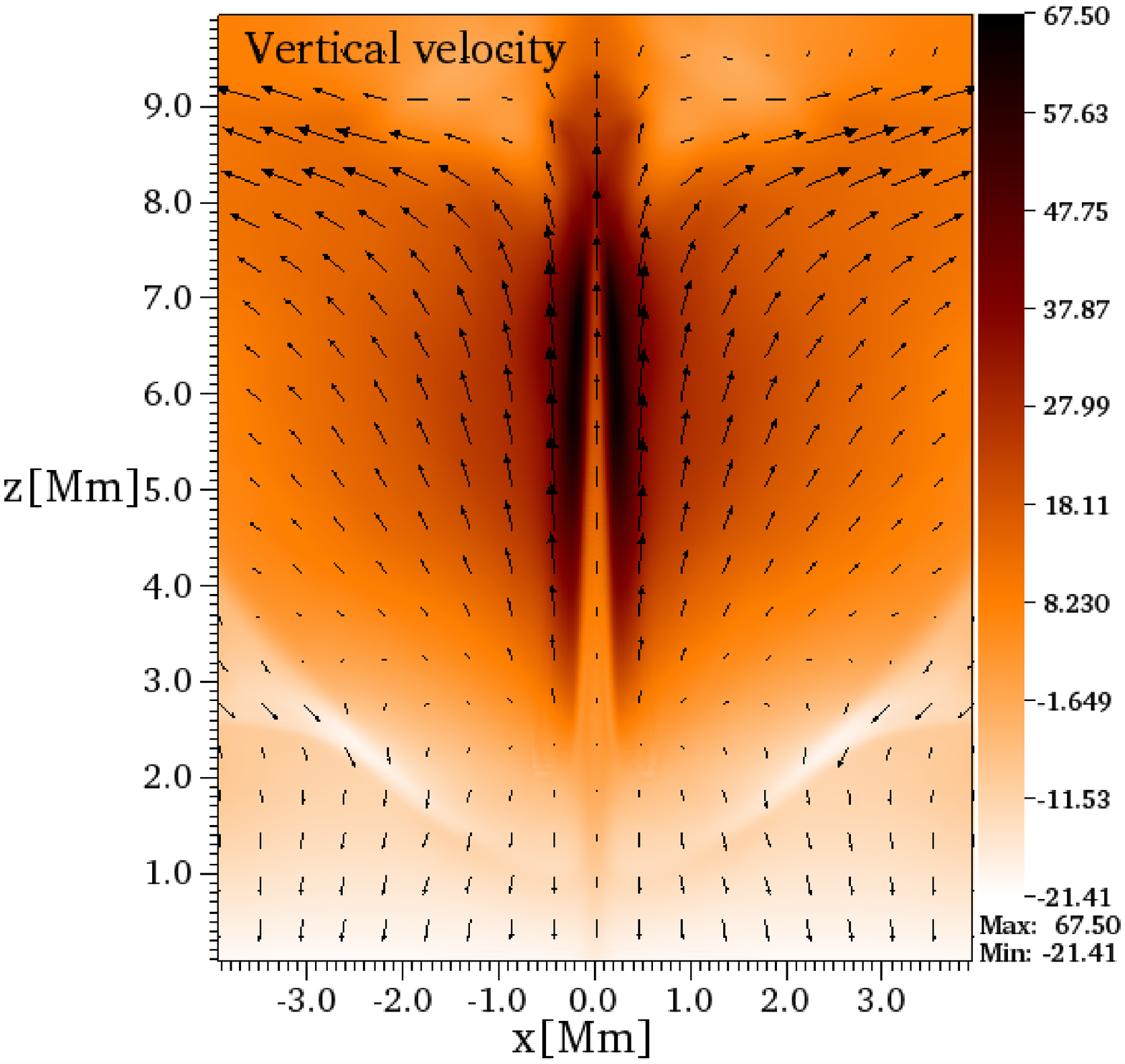}
\includegraphics[width=4.25cm,height=6.5cm]{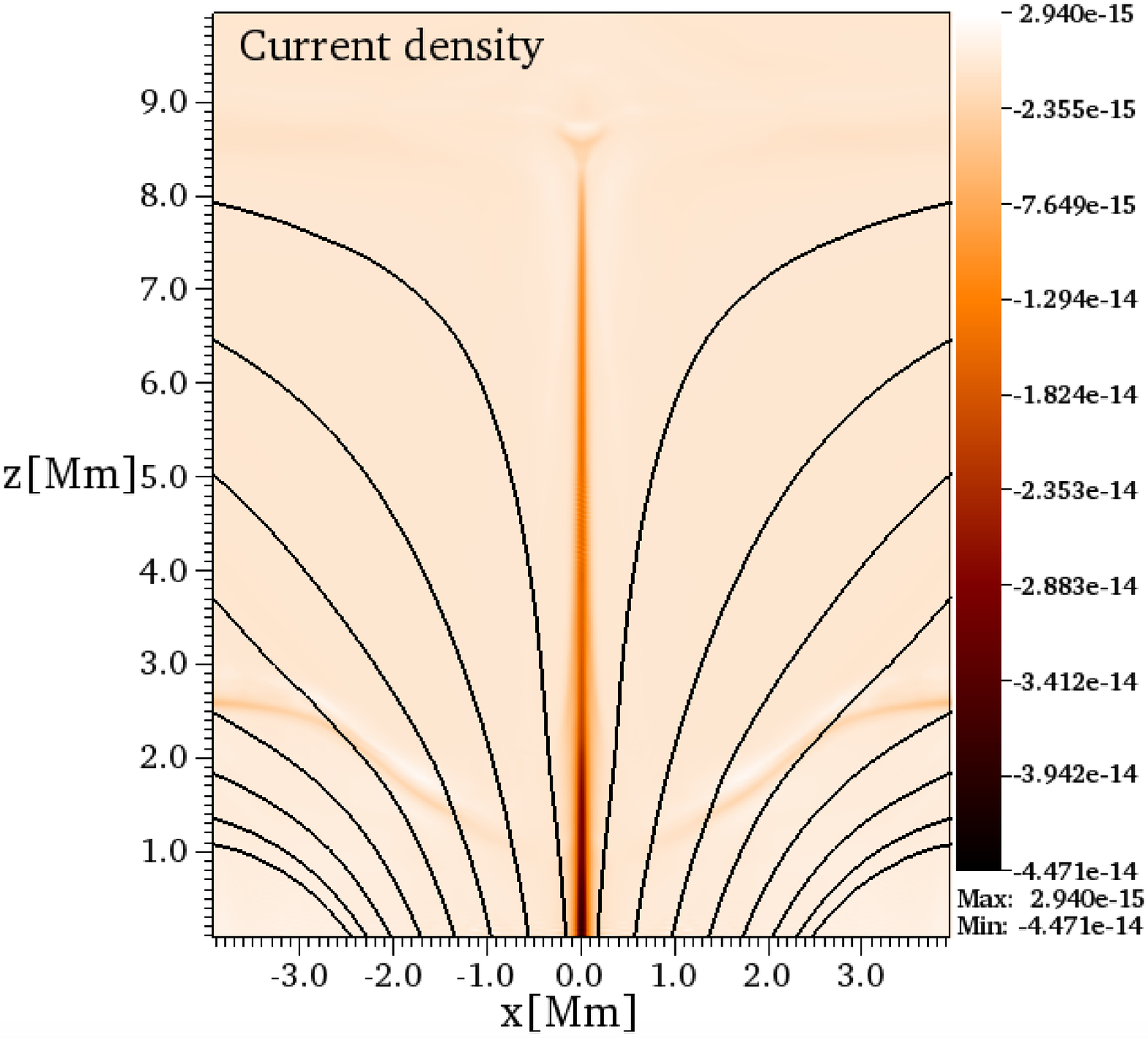}
\includegraphics[width=4.25cm,height=6.5cm]{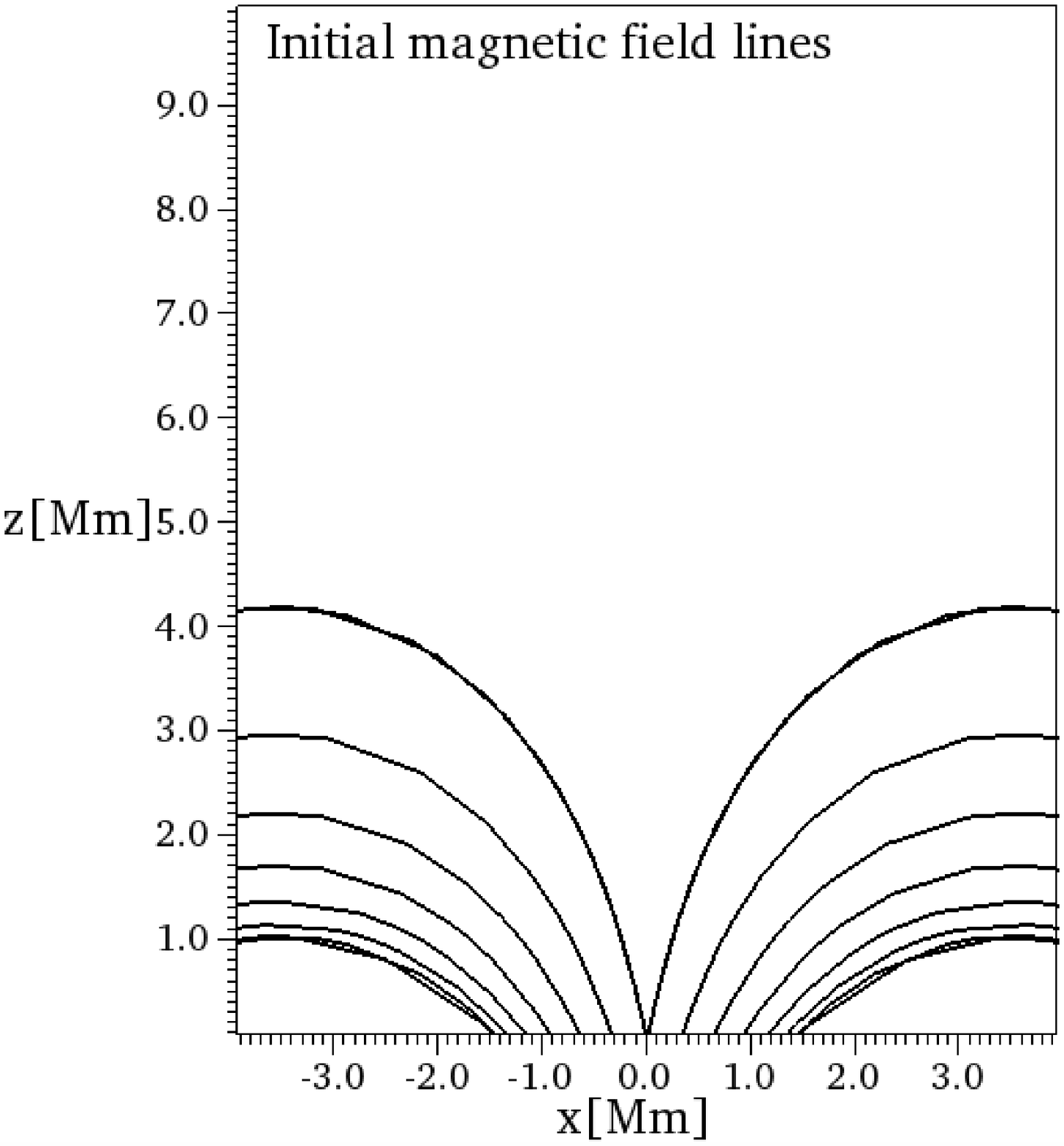}
\includegraphics[width=4.25cm,height=6.5cm]{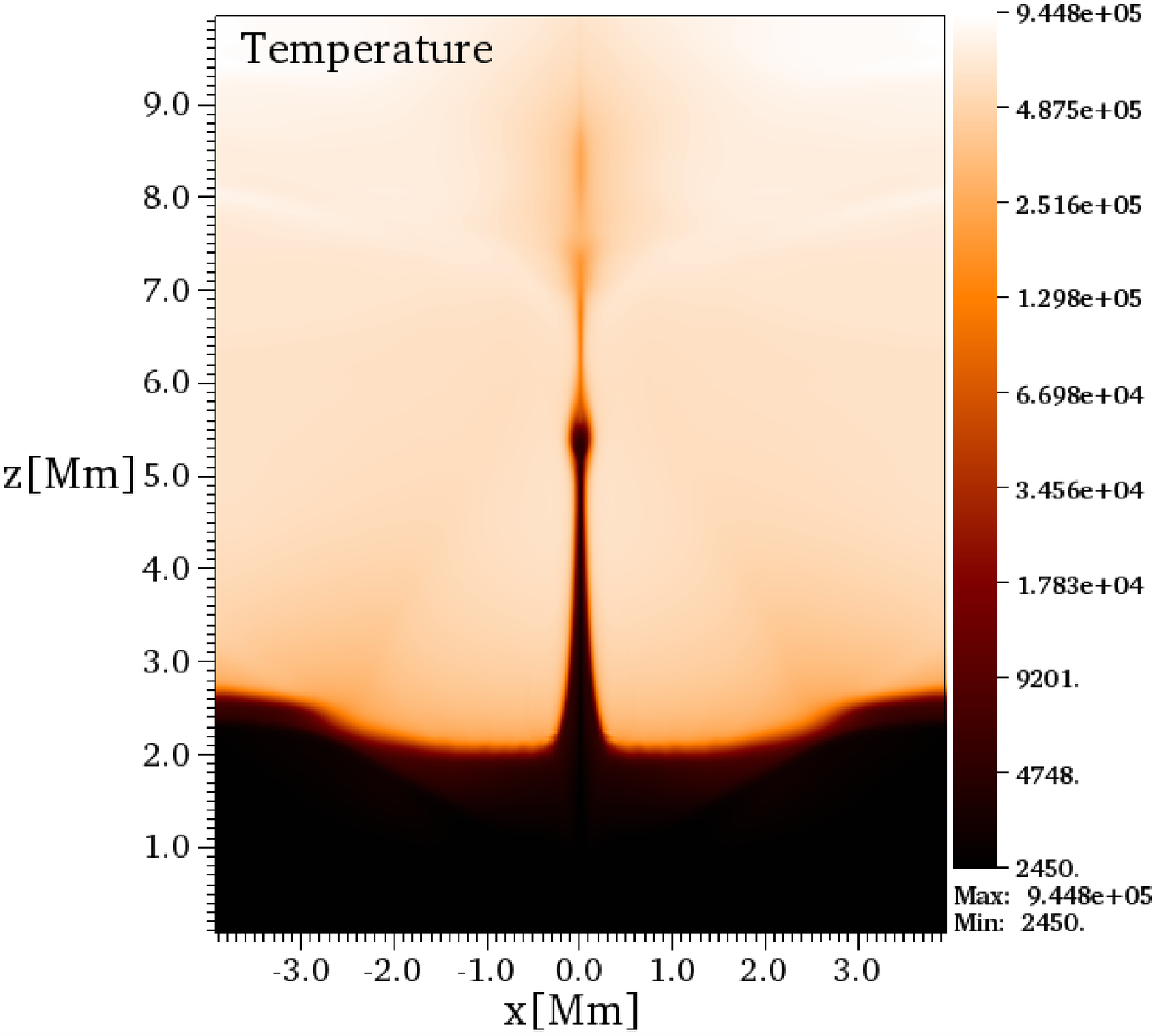}
\includegraphics[width=4.25cm,height=6.5cm]{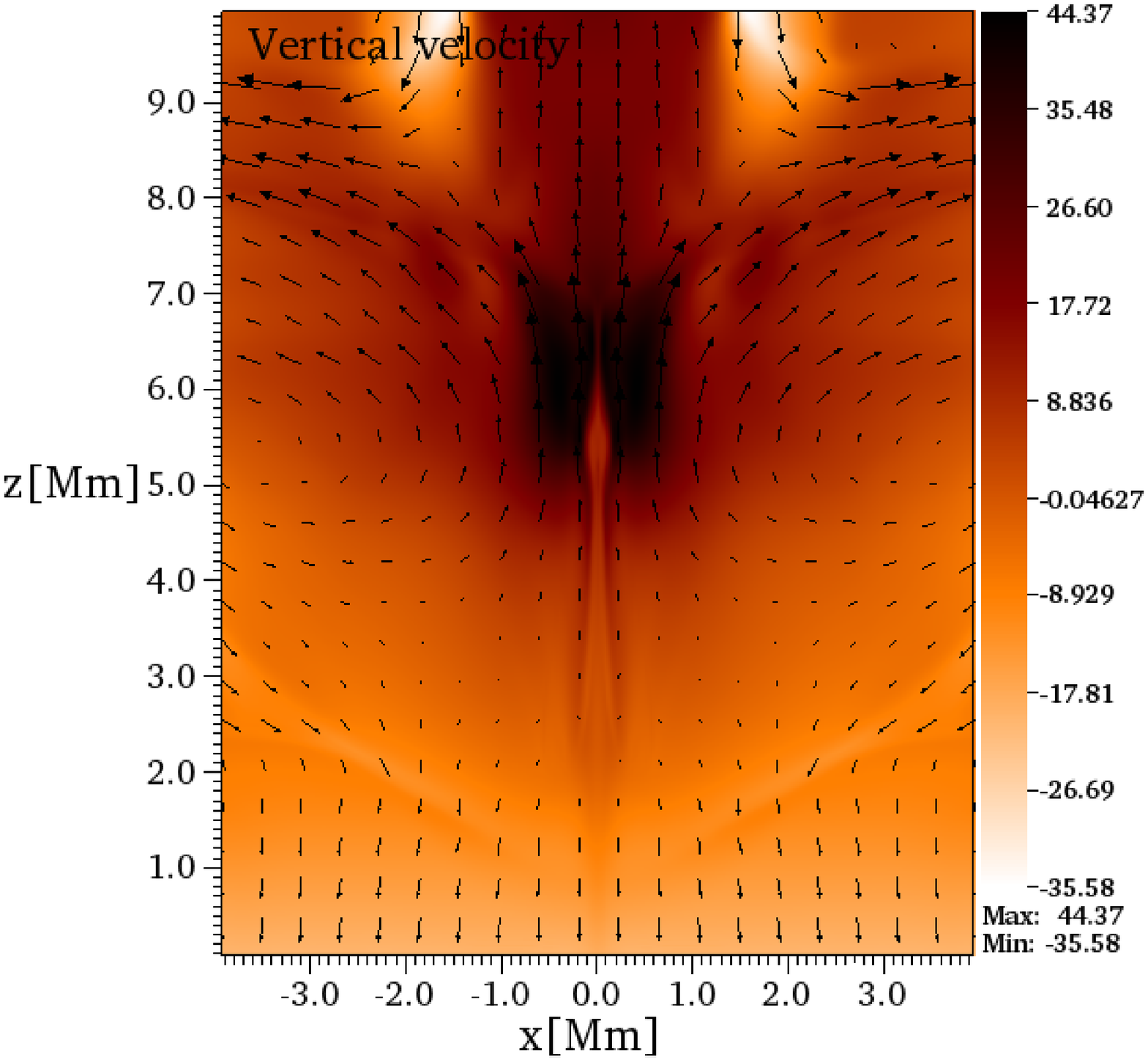}
\includegraphics[width=4.25cm,height=6.5cm]{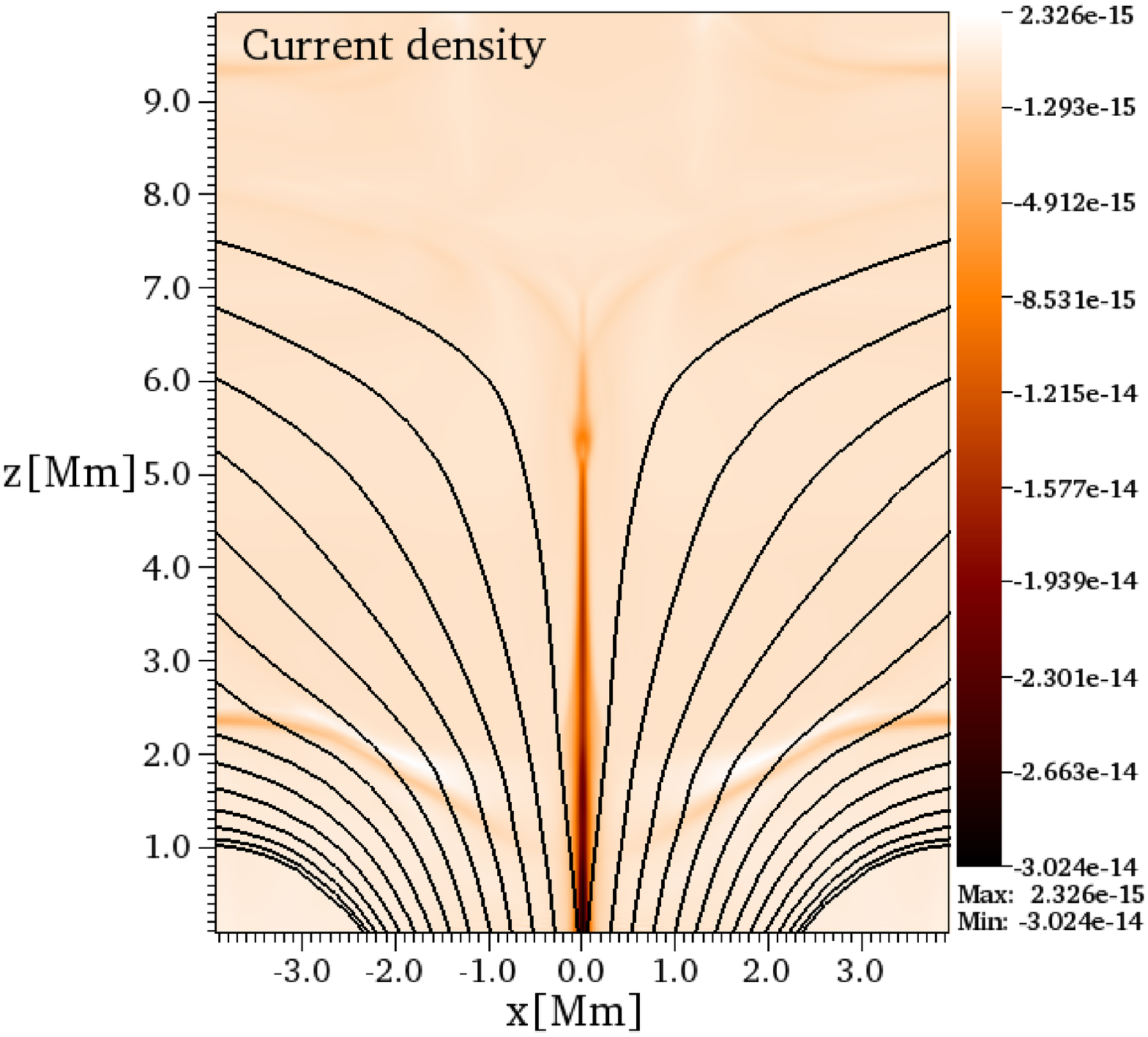}
\includegraphics[width=4.25cm,height=6.5cm]{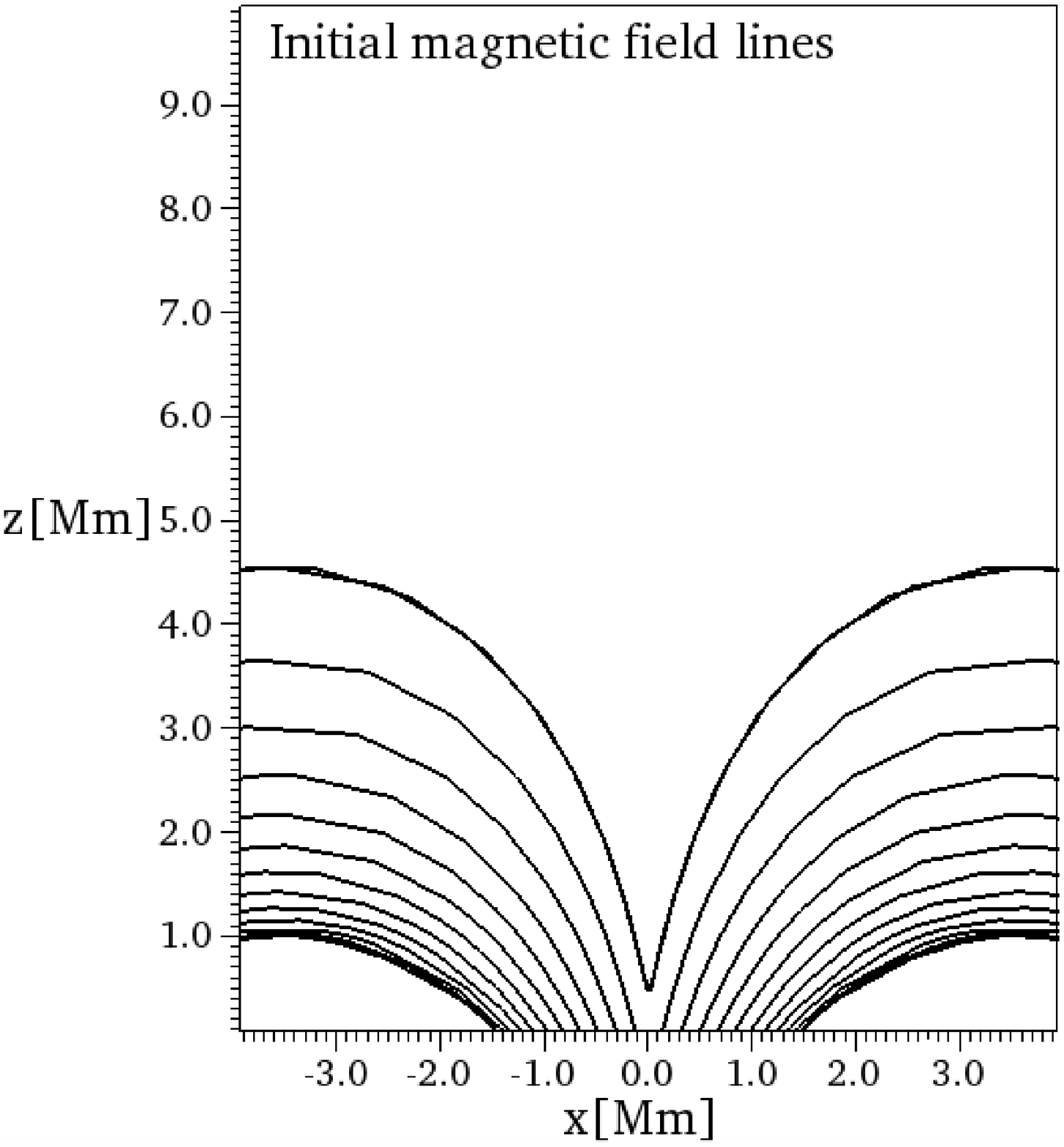}
\includegraphics[width=4.5cm,height=6.5cm]{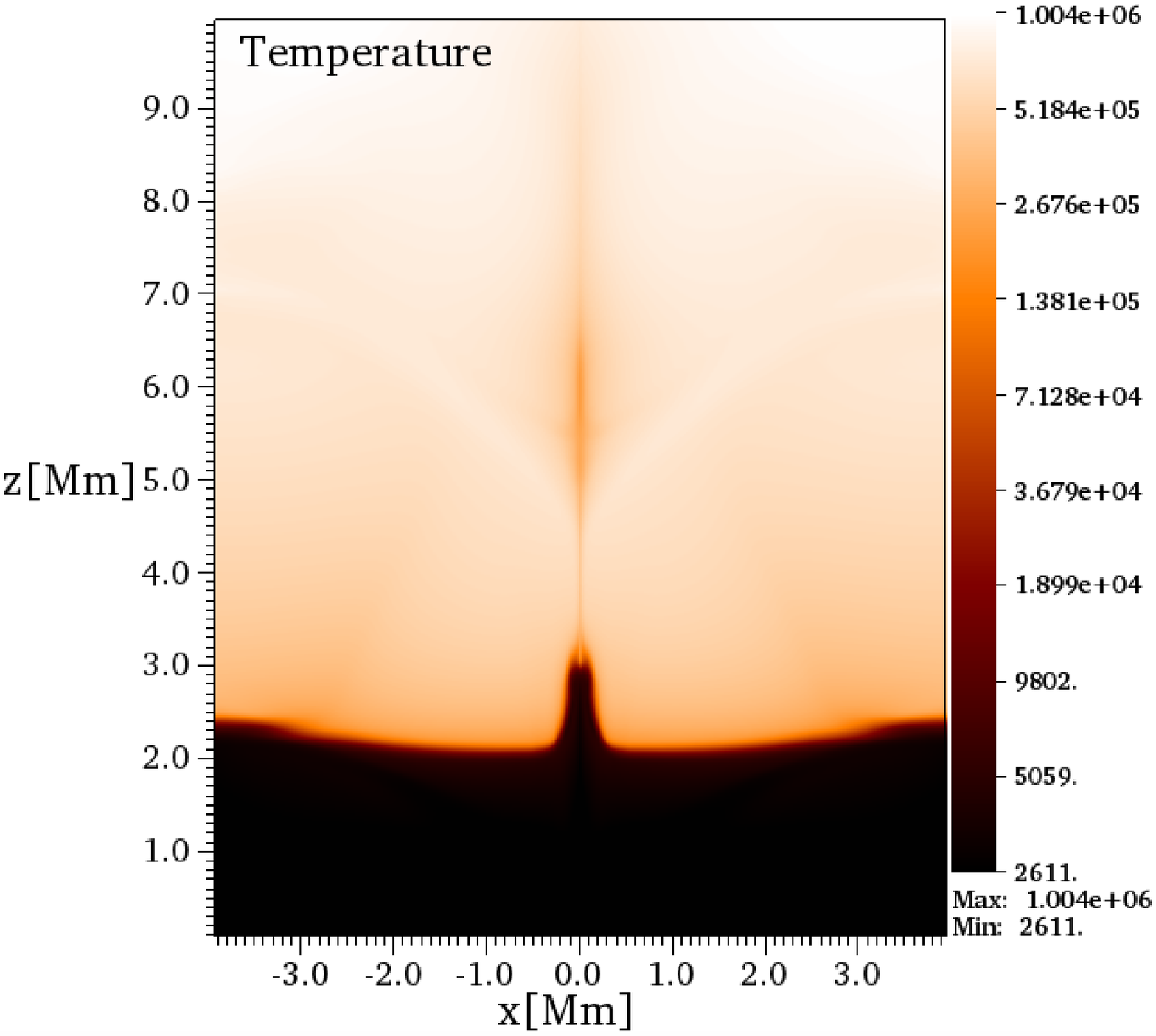}
\includegraphics[width=4.25cm,height=6.5cm]{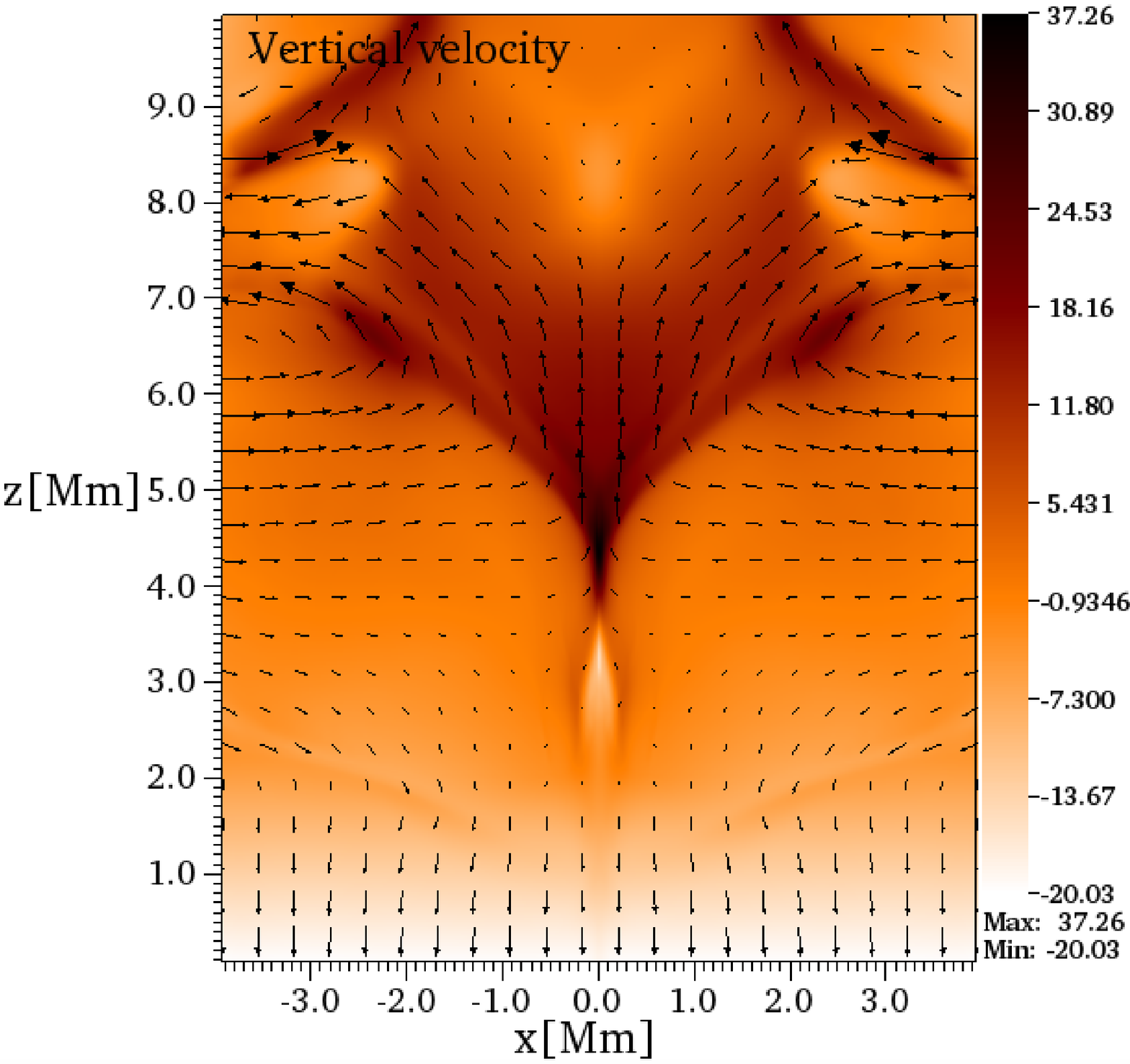}
\includegraphics[width=4.25cm,height=6.5cm]{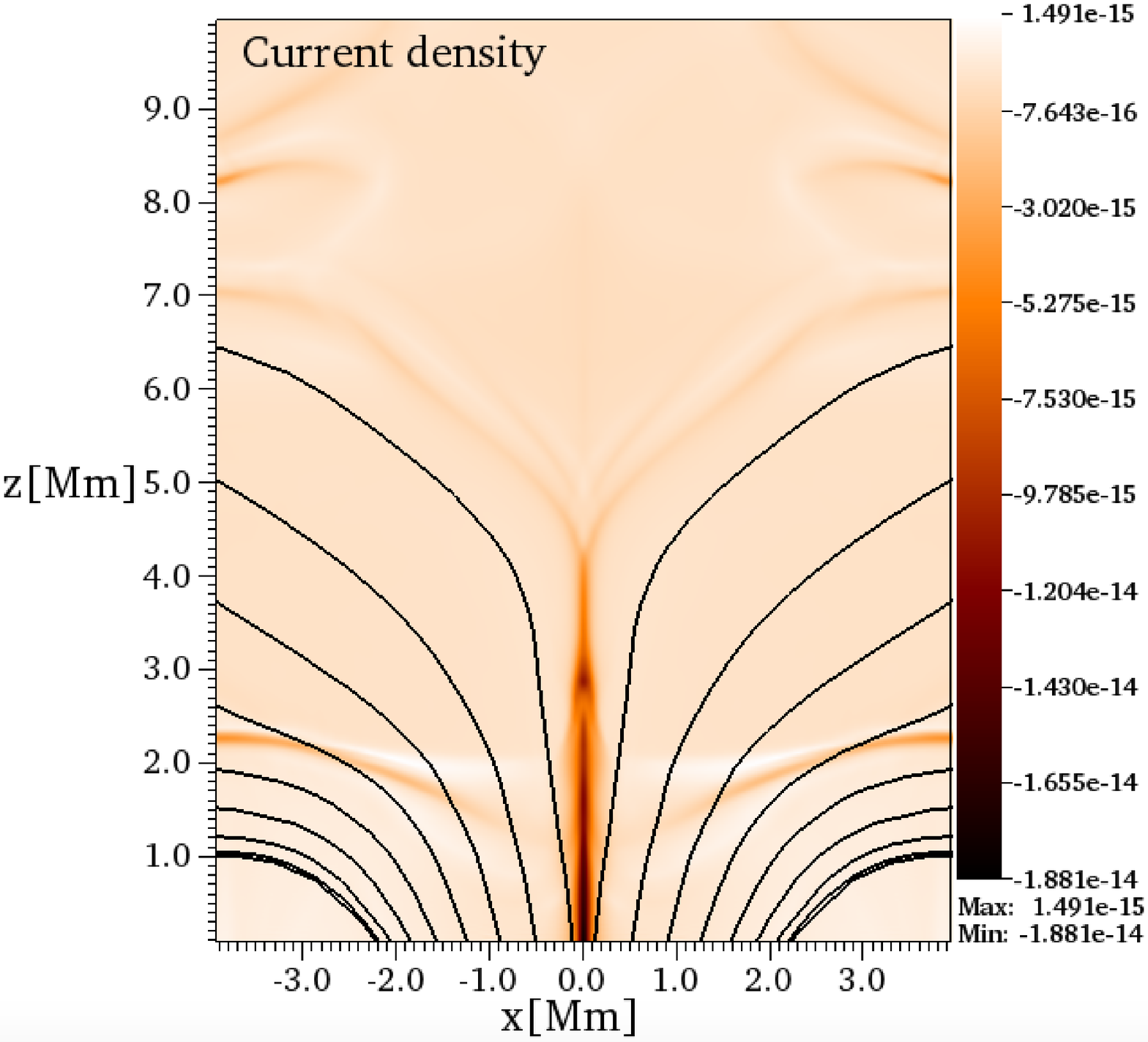}
\includegraphics[width=4.25cm,height=6.5cm]{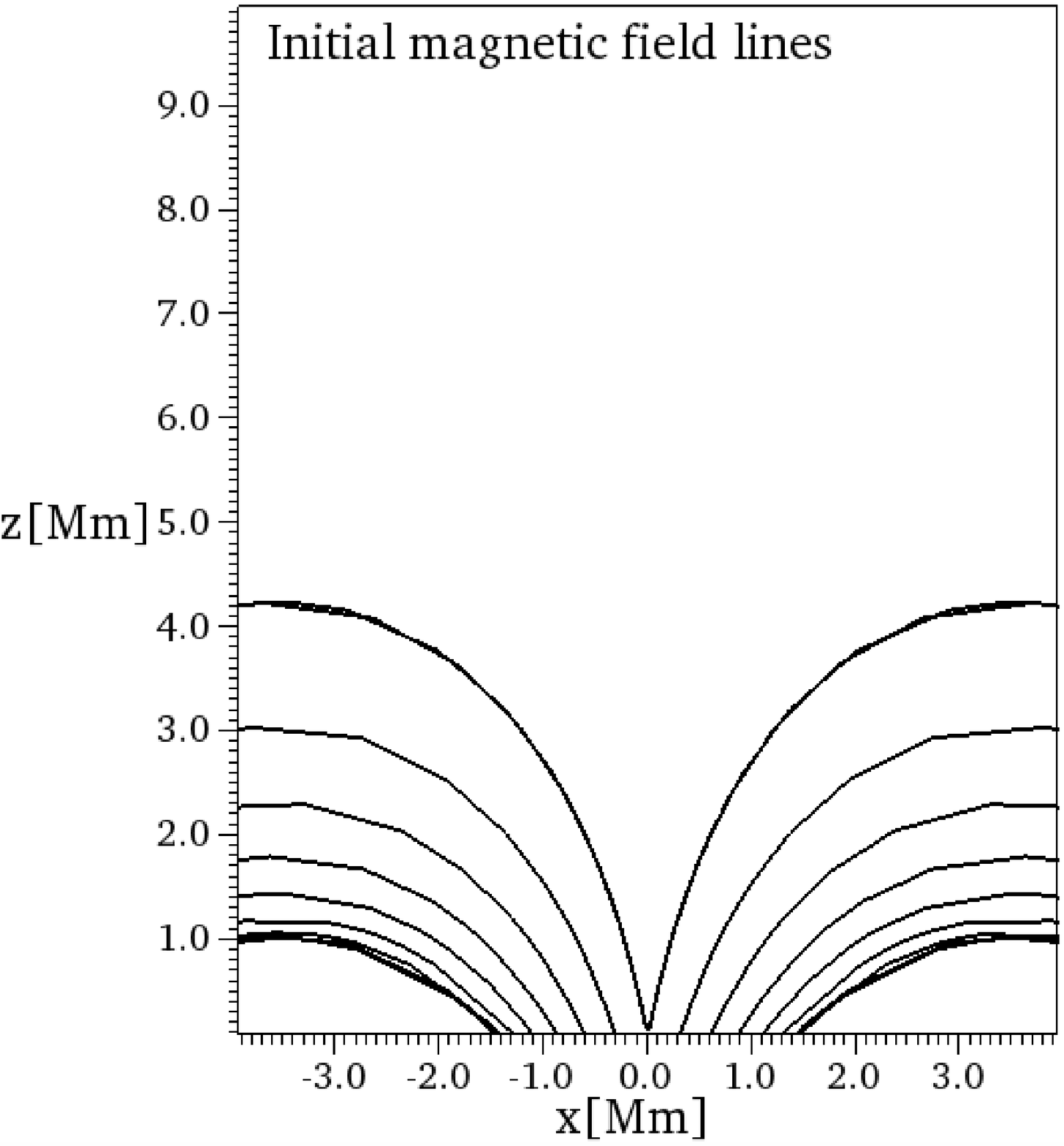}
\caption{\label{fig:caseB}  From left to right we show snapshots of  i) logarithm of the temperature in Kelvin, ii) the vertical component of the velocity ($v_z$ km/s); the arrows show the velocity field distribution, iii) the $y$ component of the current density $J_y$(A/m$^2$) at $t=180$ s,  iv) finally on the extreme right we show the magnetic field configuration at initial time, that helps noticing the distortions of the field during the evolution. In the Top panel we show the results for Run \#1, where $B_{01}=B_{02}=40$ G. In the Middle panel we present the results for Run \# 3, where  $B_{01}=B_{02}=30$ G. Finally in the Bottom we show the results for Run \# 6, where  $B_{01}=B_{02}=20$. In all the cases $l_0=3.5$Mm.}
\end{figure*}

\begin{figure}
\centering
\includegraphics[width=4.25cm,height=5.25cm]{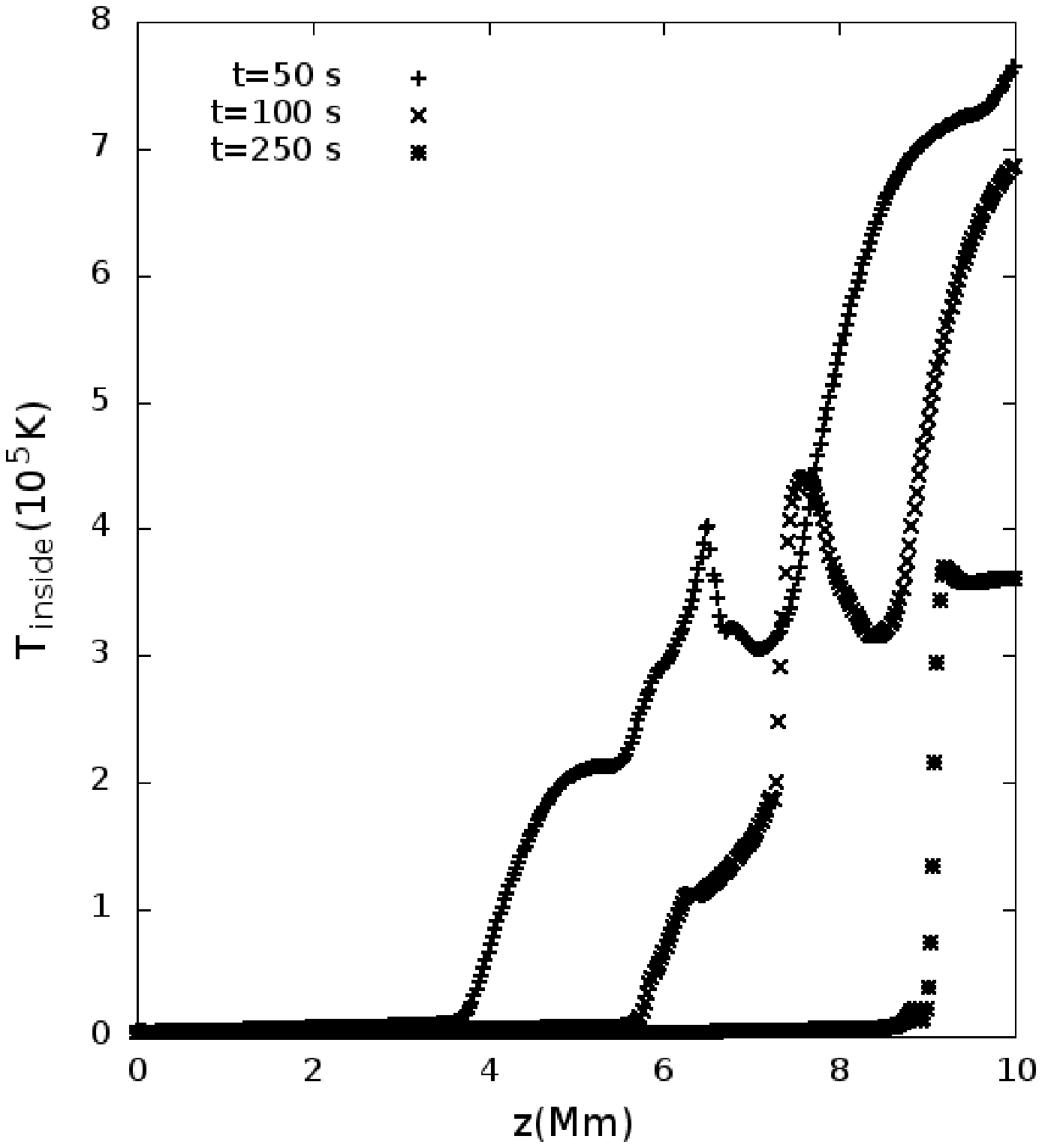}
\includegraphics[width=4.25cm,height=5.25cm]{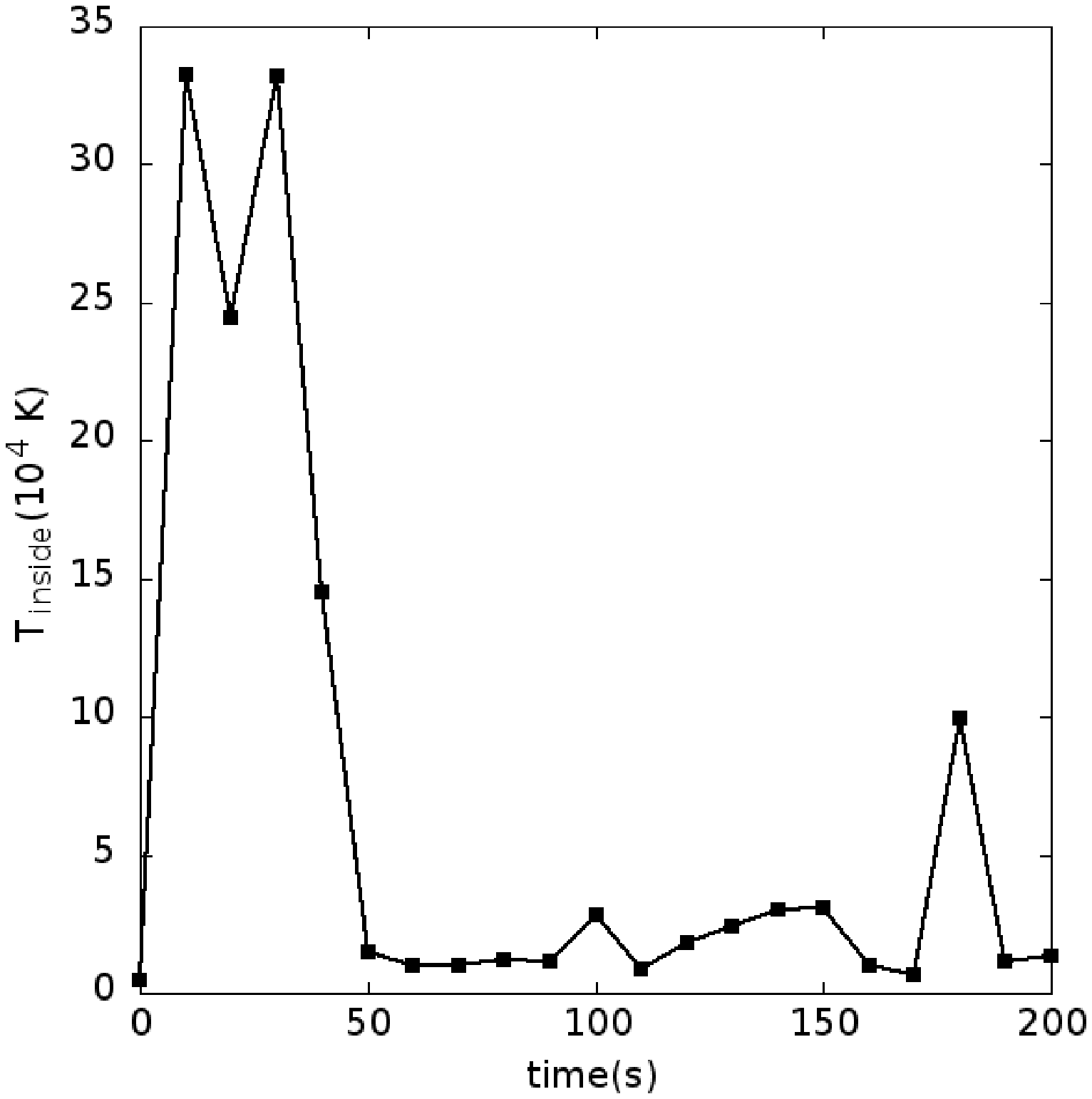}
\includegraphics[width=4.25cm,height=5.25cm]{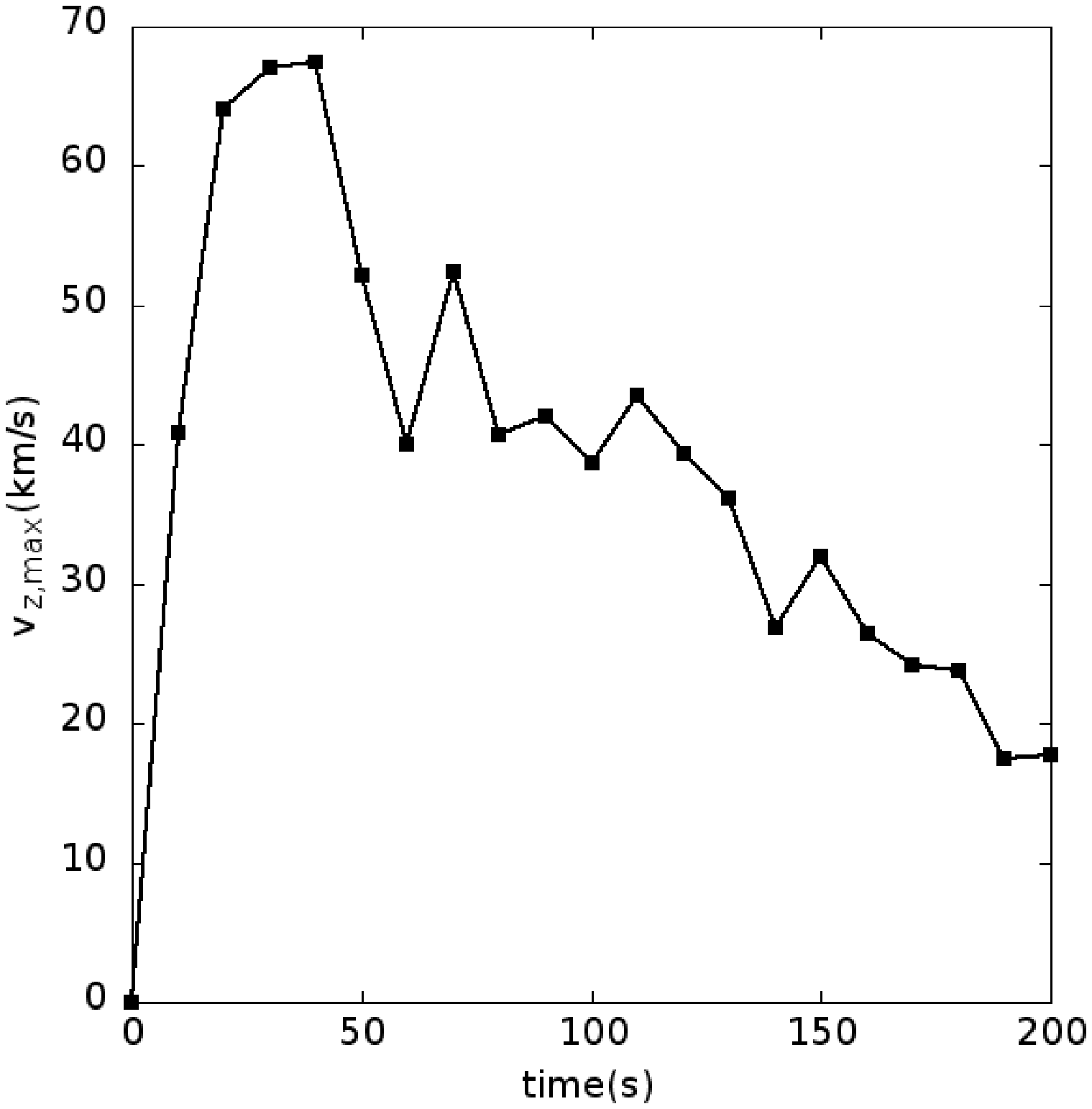}
\includegraphics[width=4.25cm,height=5.25cm]{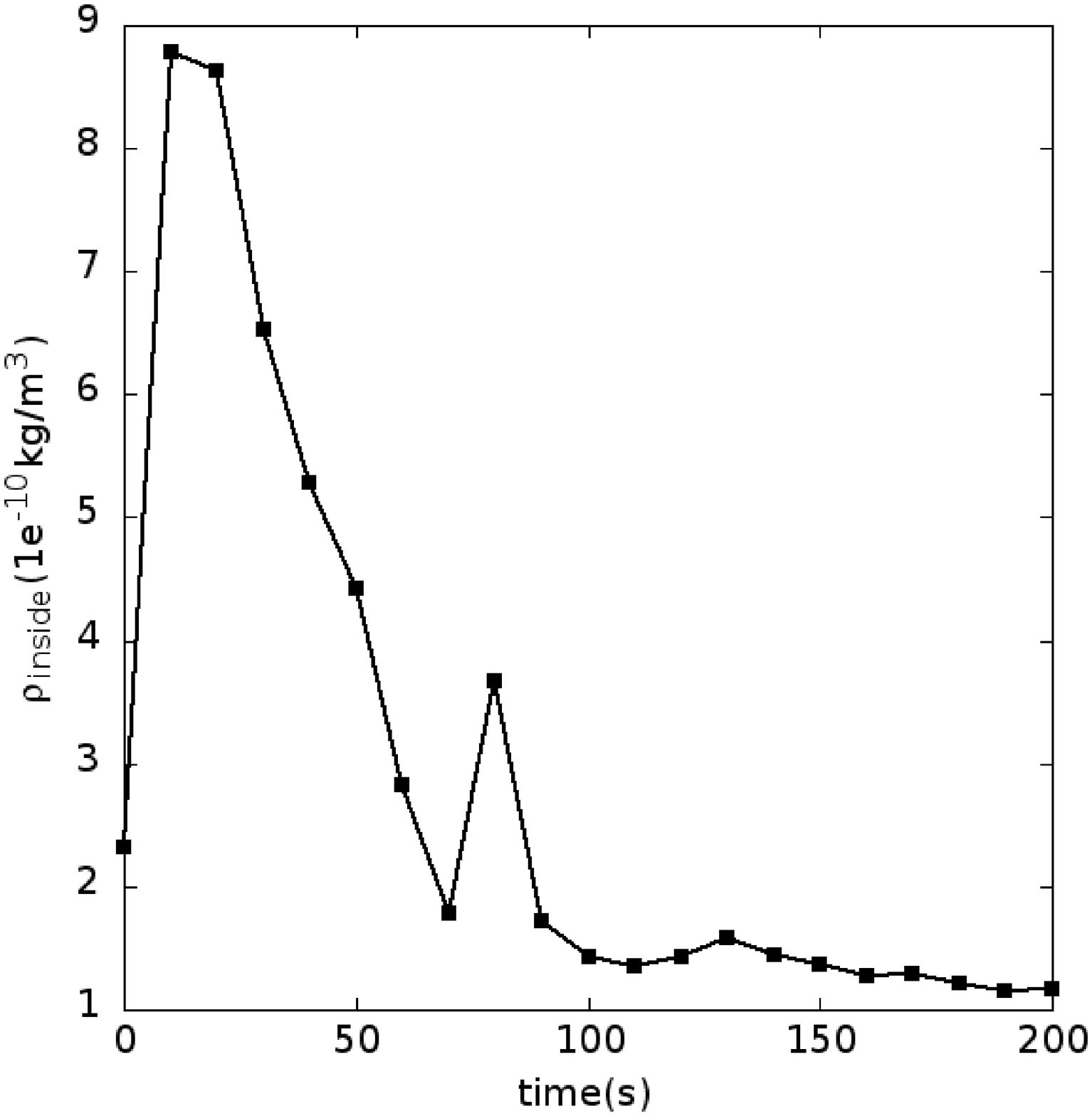}\\
\caption{\label{fig:vars_vs_time} Here we illustrate the way we measure the properties of the jet using diagnostics of the case $B_{01}=B_{02}=40$ G and $l_0=3.5$ Mm. On the top-left we show snapshots of the plasma temperature at $t=50,100,250$ s along the $z$-axis. The plasma moving upwards during the evolution is cold with respect to the temperature of the corona and therefore the temperature shows a minimum where the front of the jet is developing. We say the Temperature of the jet is that of the minimum and changes with time and position. We define the temperature of the jet $T_{inside}$ as that of the minimum at every time, independently of the location of such minimum in space and the result appears in the top-right panel. With a similar procedure we measure $v_z$ of the jet as the maximum along the $z$-axis (remember that in the case of $T$ it was a minimum) and plot this quantity in time $v_{z,max}$ that we show in the bottom-left panel. Finally we also estimate the jet density $\rho_{inside}$ with a similar procedure and show the result in the bottom-right.}
\end{figure}

For the first case show the results for the case $B_{01}=B_{02}$ for three values of magnetic strengths $20,~30,~40$ G and $l_0=~2.5,~3.0,~3.5$ Mm. Representative simulations for this case are shown in Fig. \ref{fig:caseB} for $l_0=3.5$ Mm. For Run  \#1 correspond to the typical formation of a jet, with a special feature at the top with a bulb related to a Kelvin-Helmholtz type of instability which is contained due to the presence of the magnetic field. This jet reaches a height of  9 Mm with a maximum speed $v_{z,max}\approx 34$ km/s at time $t=180$ s. After this time the jet starts falling down until it disperses away by time $t\approx 400$ s. In Fig. \ref{fig:vars_vs_time} we show $T_{inside}$, $\rho_{inside}$ and $v_{z,max}$ as a function of time estimated inside the jet, and the method used to estimate this properties of the jet. The temperature during the evolution is a useful scalar at determining the location of the jet, because it shows a minimum precisely located where the head of the jet is and is used to estimate the maximum height achieved.

We also show in Fig. \ref{fig:caseB} the $y$ component of current density $J_y$,  which is the most significant component and shows the formation of an elongated structure representing the reconnection process. At the middle panel of Fig. \ref{fig:caseB} we show the results of Run \#3 (the magnetic field strength is 20 G). Due to smaller magnetic field strength the resulting jet reaches a height of 5.5 Mm with a maximum vertical velocity $v_{z,max}\approx 32.5$ km/s at time $t=180$ s. Finally at the lower panel of Fig. \ref{fig:caseB} we show Run \#6, in this case the effect of a weaker magnetic field is seen in the appearance of a small jet that reaches a height of 3 Mm with a maximum speed $v_{z,max}\approx 21.32$ km/s at time $t=180$ s. These results indicate that the height of the jet is stronger for bigger values of $B_{01}=B_{02}$, and the sharpness of the jet is also more clear for  magnetic fields. We present the snapshots of all the cases at $t=180$ s for comparison, the maximum height and velocity are different in each case.  

The separation of the loops is also important due to the dependence of the current sheet parameters on it.  In the case of configurations with a larger separation, the plasma is accelerated rapidly, which produces diffusion and consequently the jet does not form. However in the case of closer loops, the plasma is accelerated slowly, which allows the formation of a jet later on.   According to our results, for the parameters we analyzed, the most effective separation between the loops to trigger a jet is $l_0=3.5$ Mm.

\subsection{Non-symmetric configurations}
\label{sub_sec:non-symmetric_loop_results}

\begin{figure*}
\centering
\includegraphics[width=4.25cm,height=6.5cm]{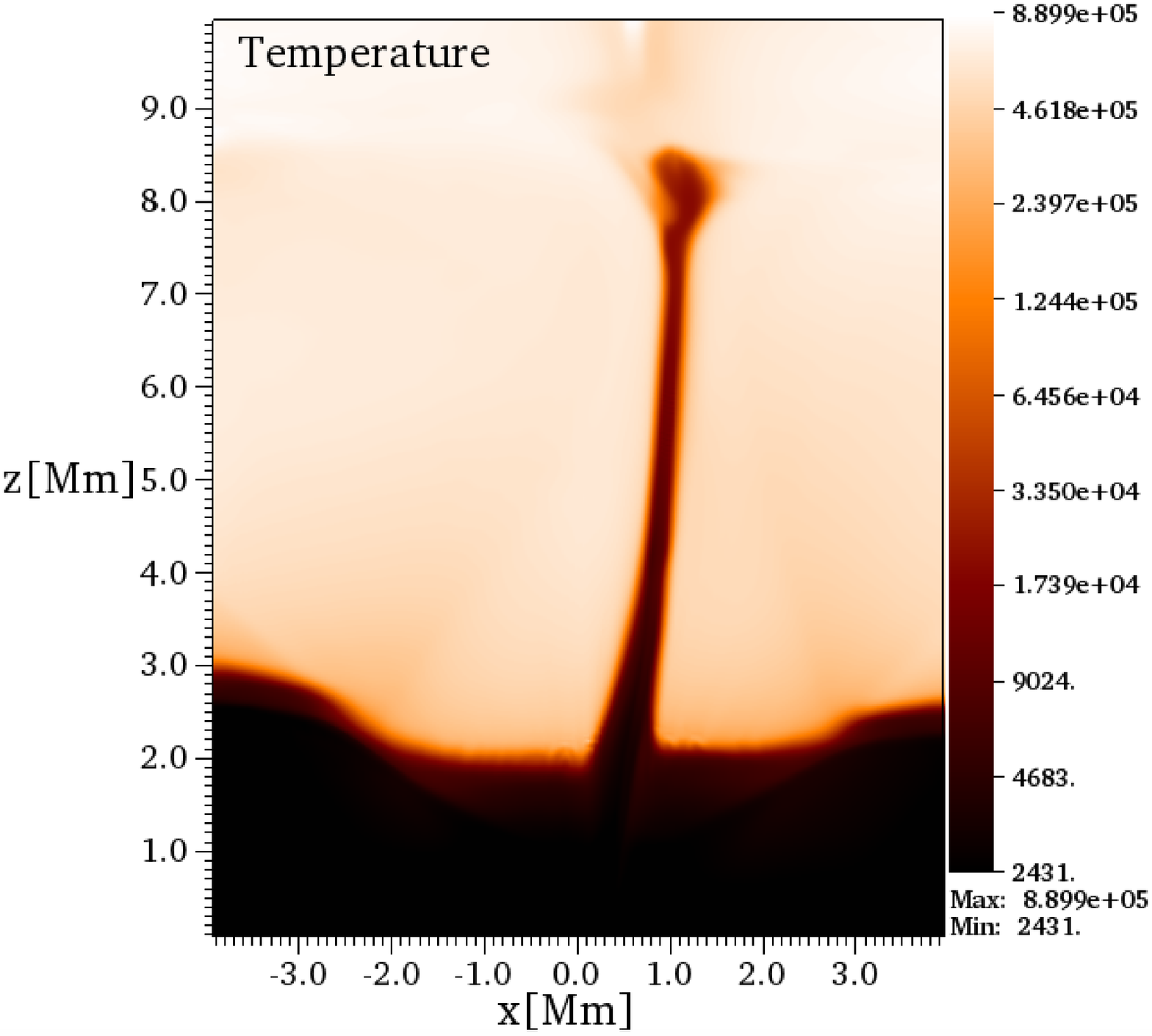}
\includegraphics[width=4.25cm,height=6.5cm]{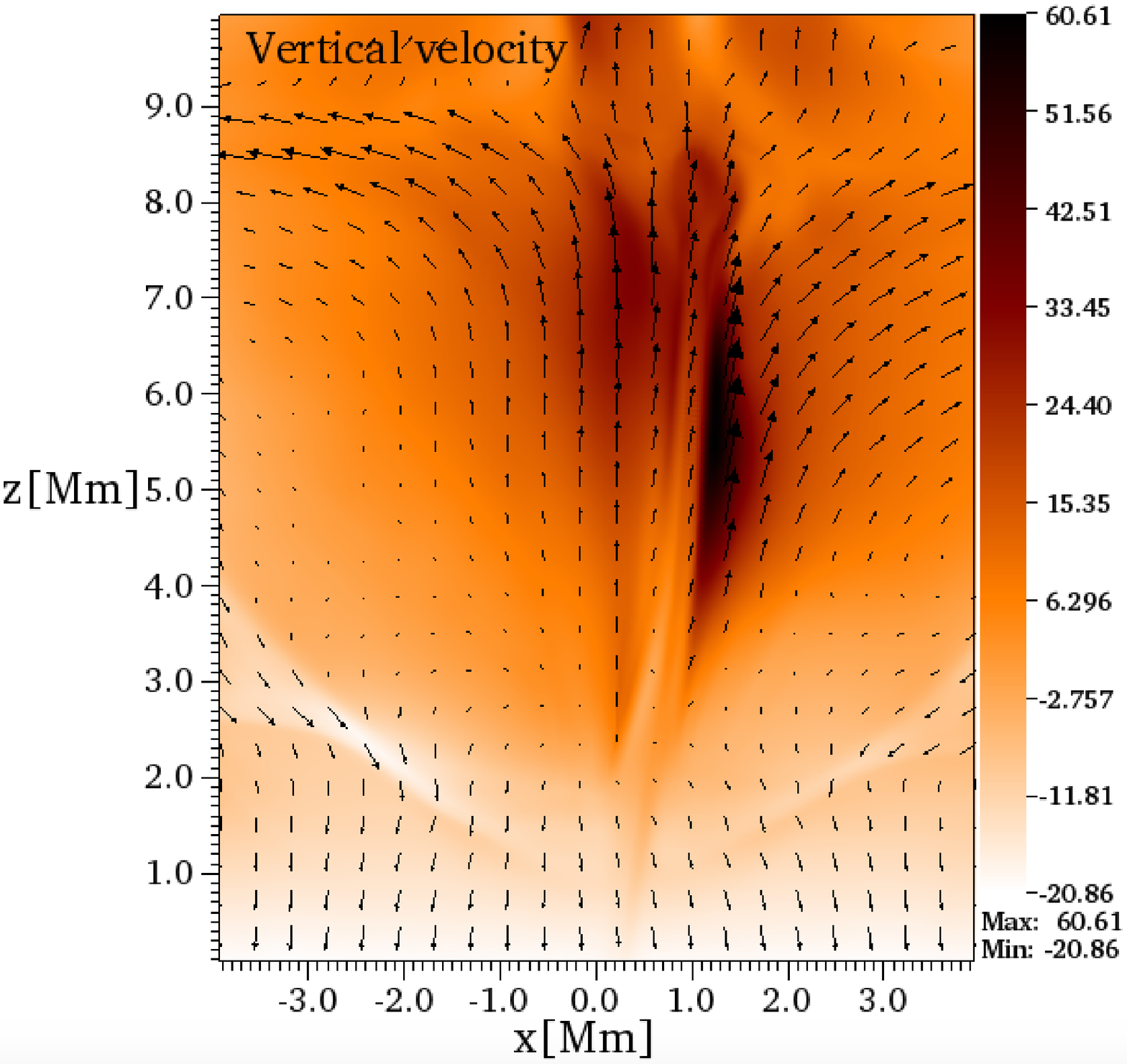}
\includegraphics[width=4.25cm,height=6.5cm]{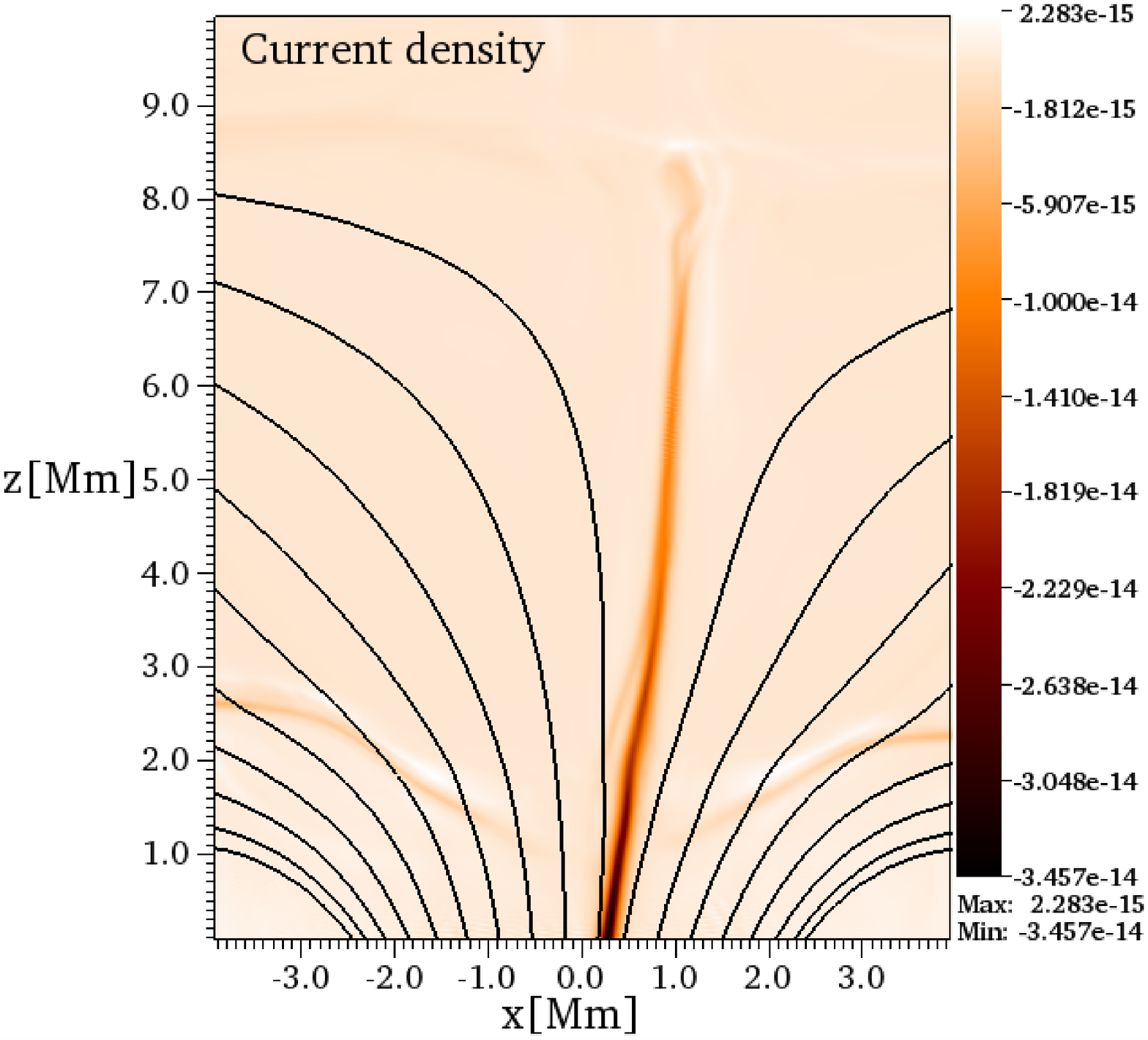}
\includegraphics[width=4.25cm,height=6.5cm]{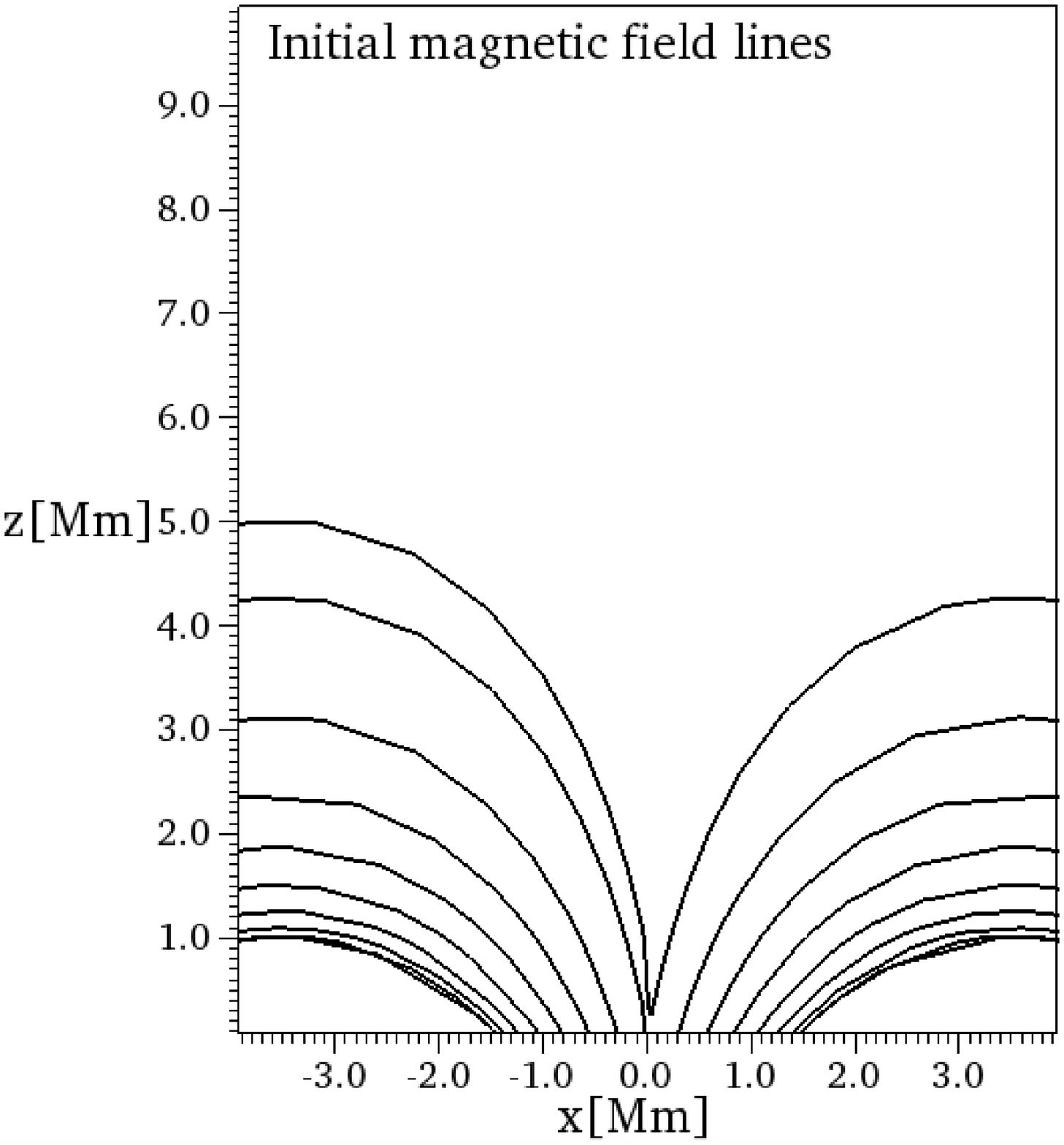}
\includegraphics[width=4.25cm,height=6.5cm]{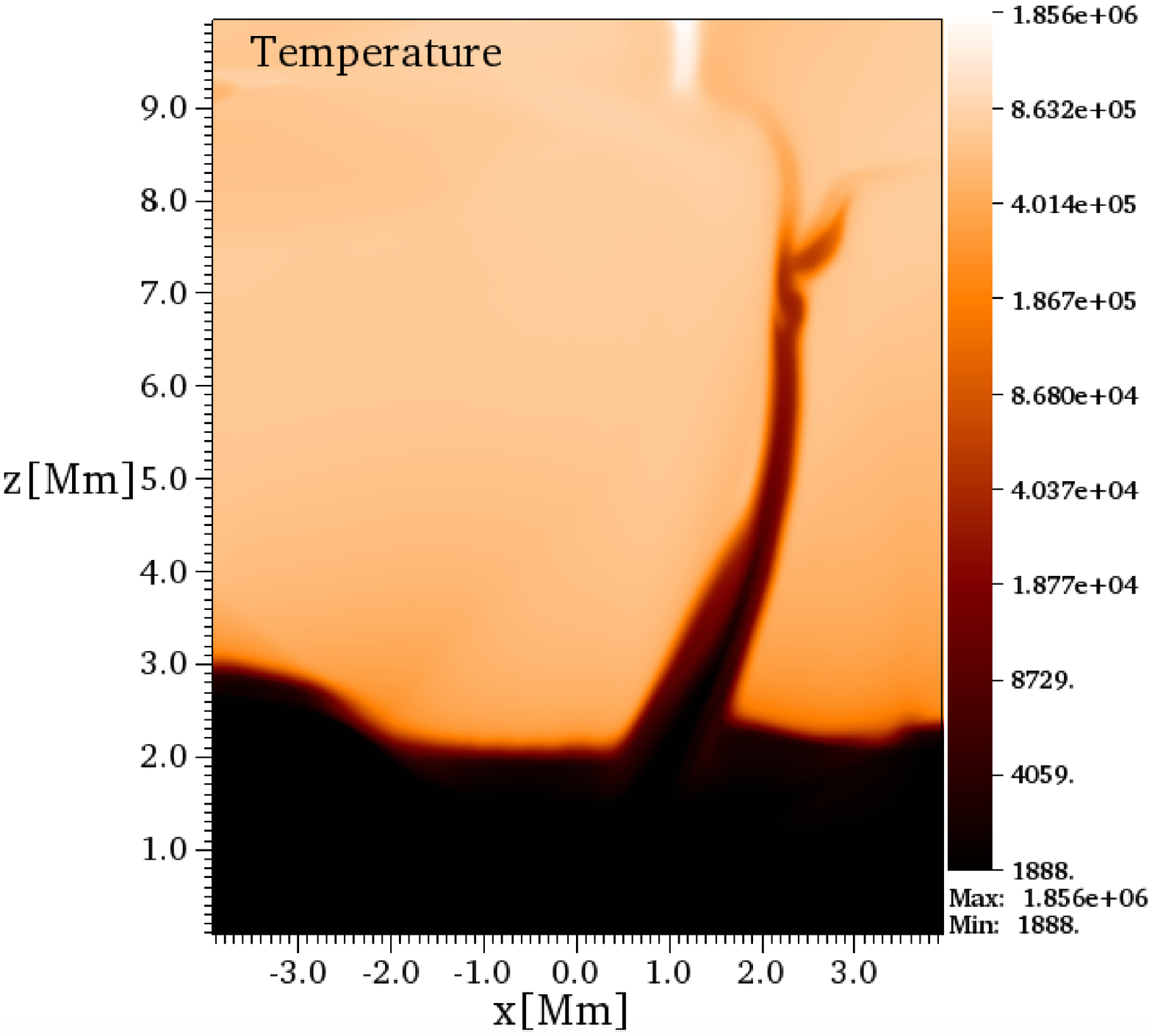}
\includegraphics[width=4.25cm,height=6.5cm]{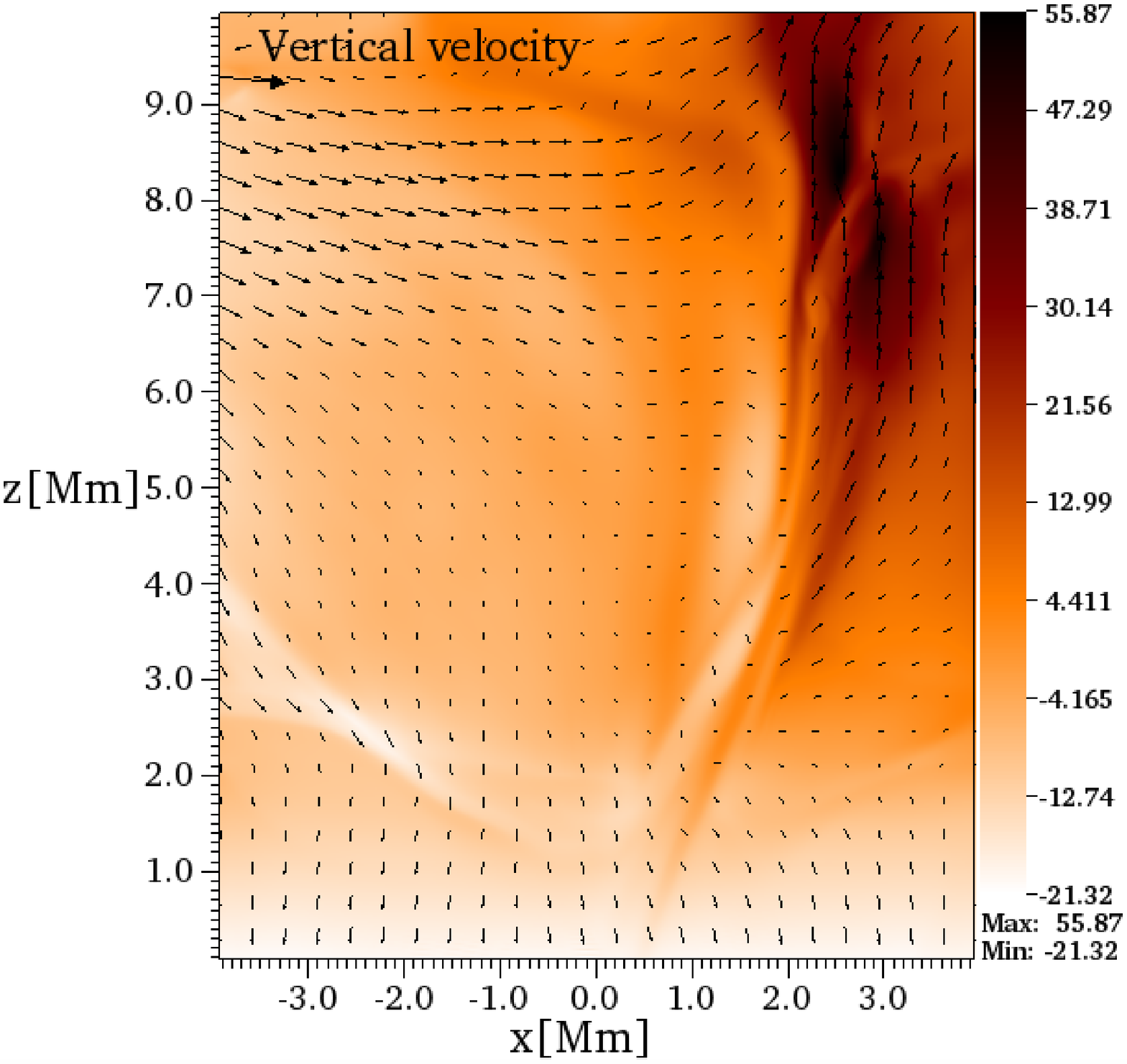}
\includegraphics[width=4.25cm,height=6.5cm]{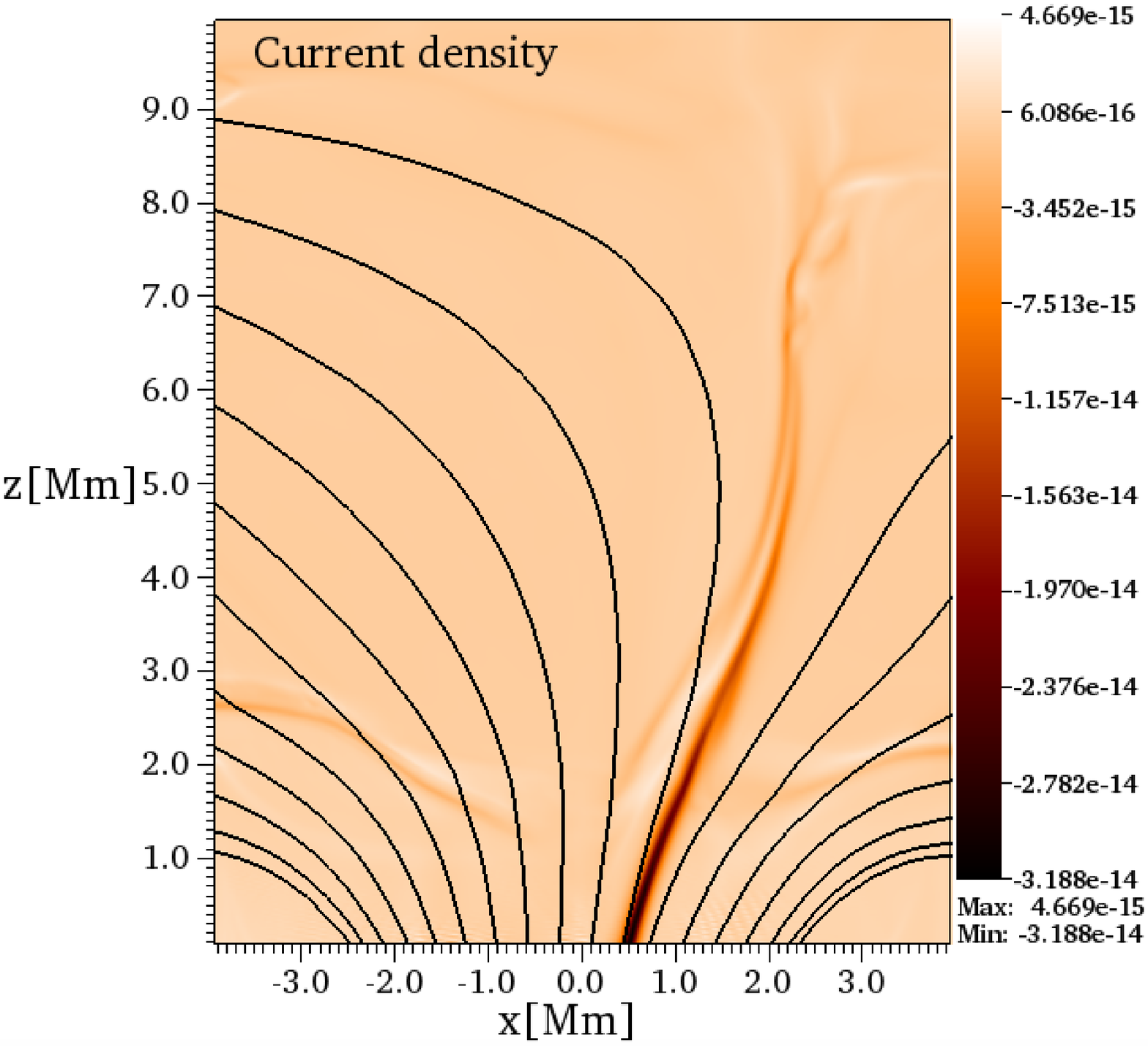}
\includegraphics[width=4.25cm,height=6.5cm]{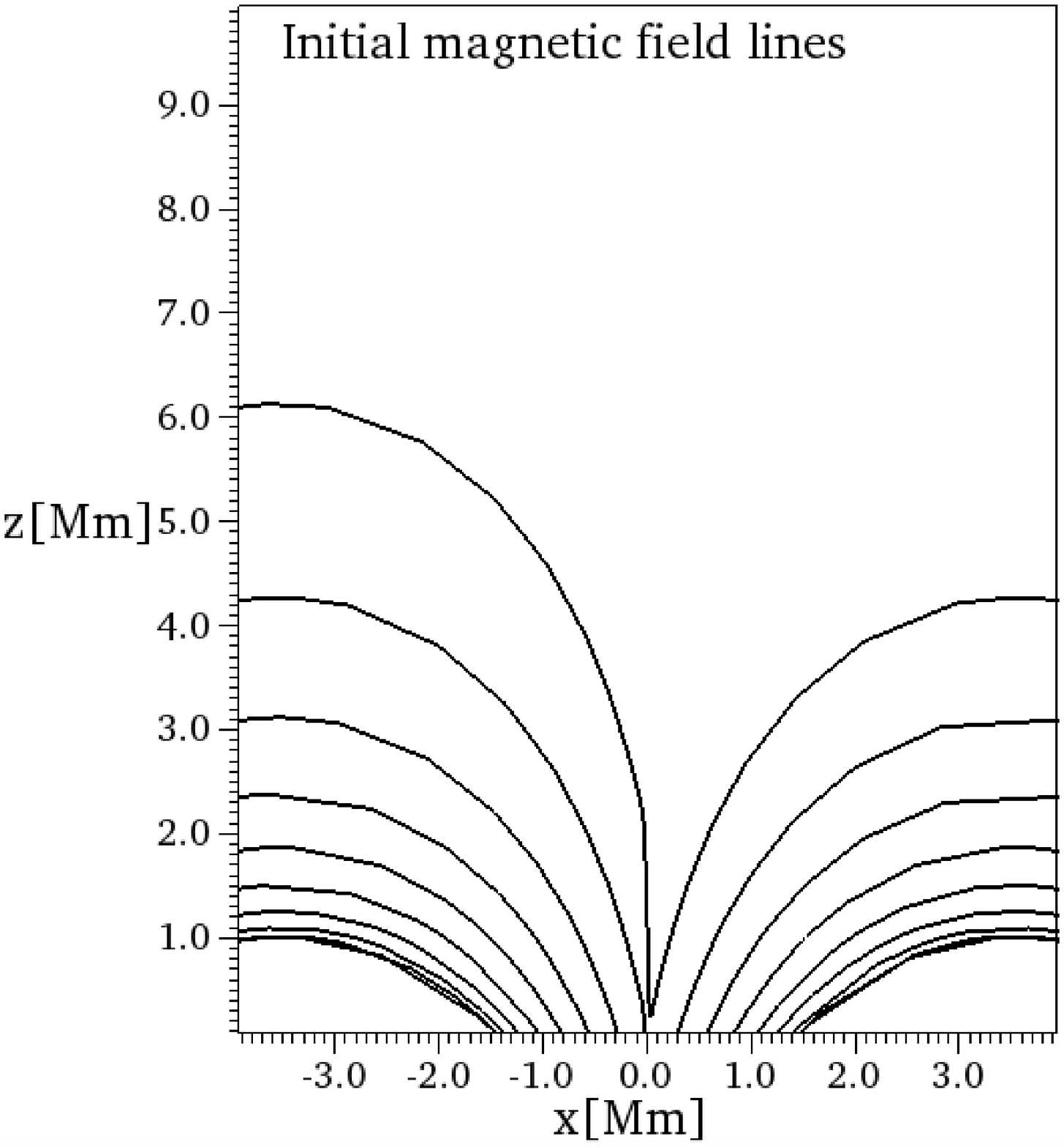}
\includegraphics[width=4.25cm,height=6.5cm]{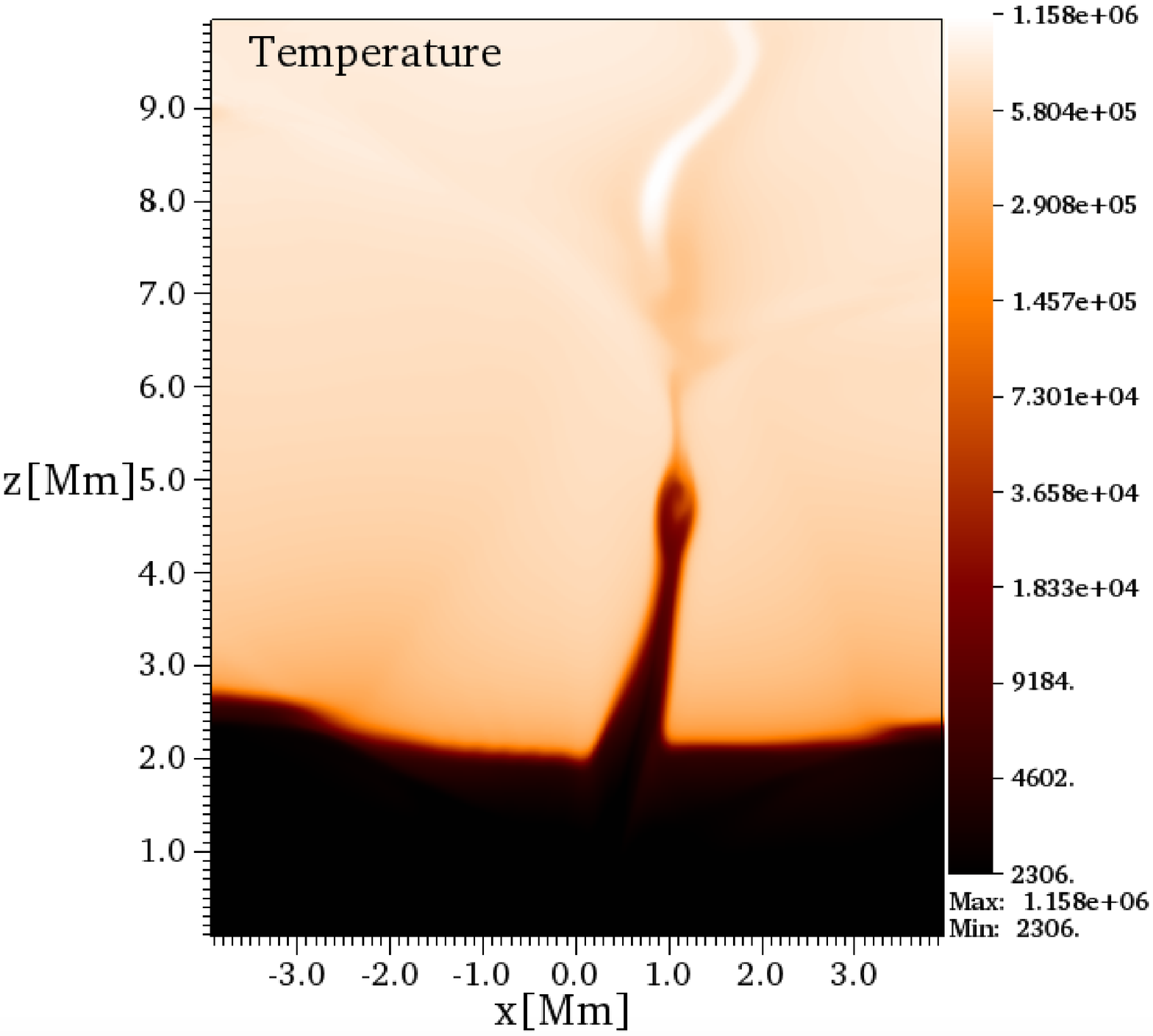}
\includegraphics[width=4.25cm,height=6.5cm]{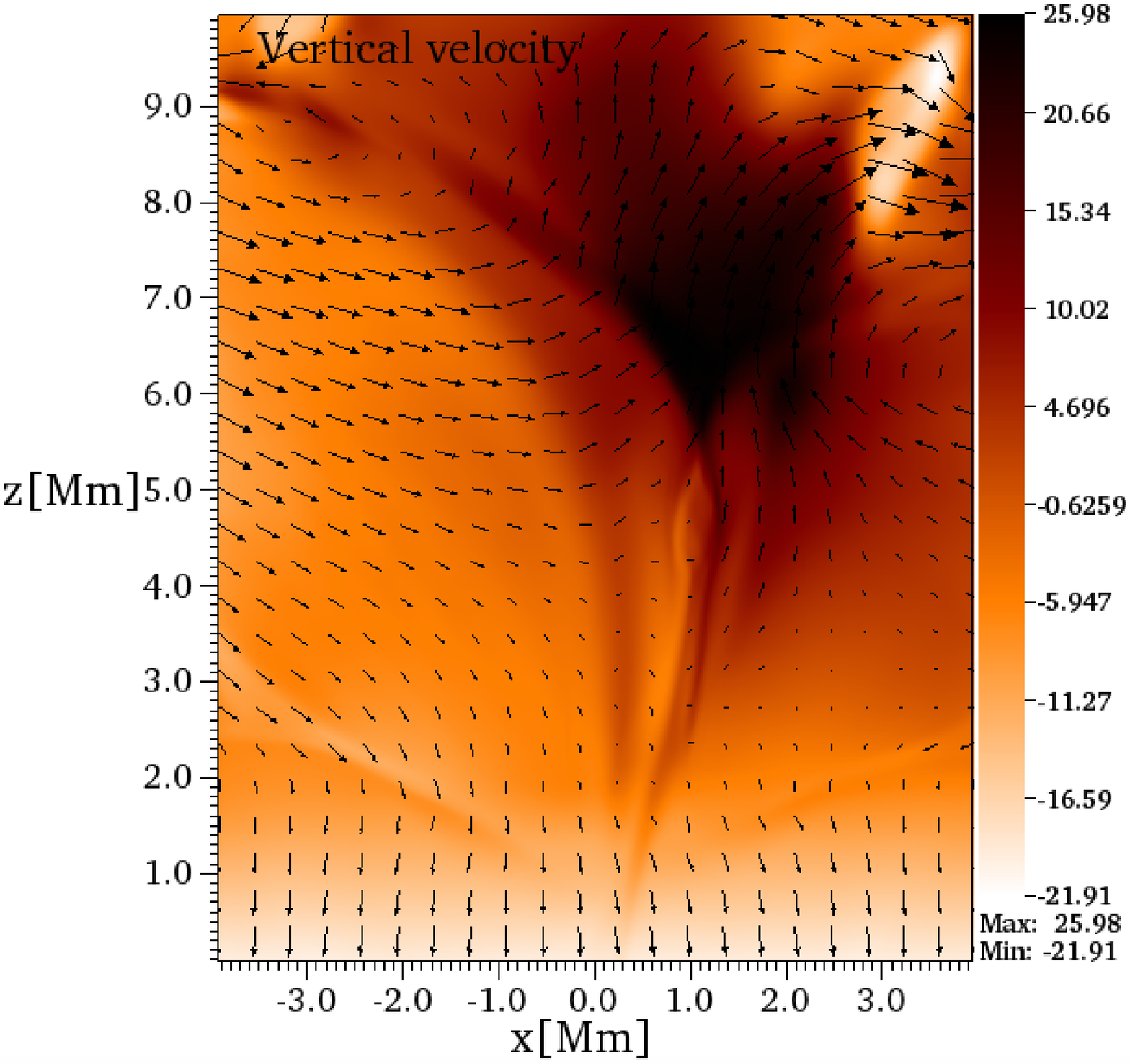}
\includegraphics[width=4.25cm,height=6.5cm]{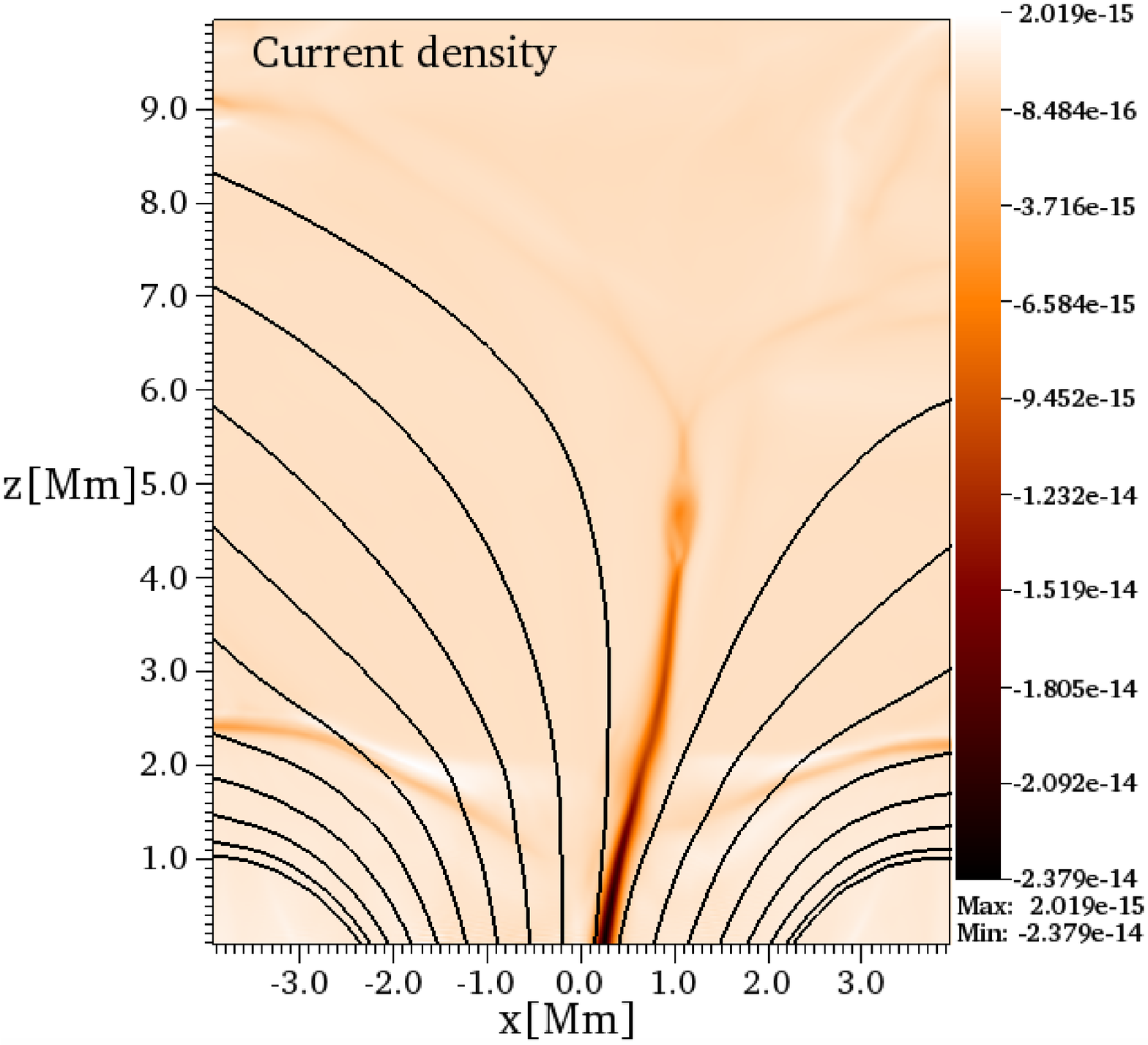}
\includegraphics[width=4.25cm,height=6.5cm]{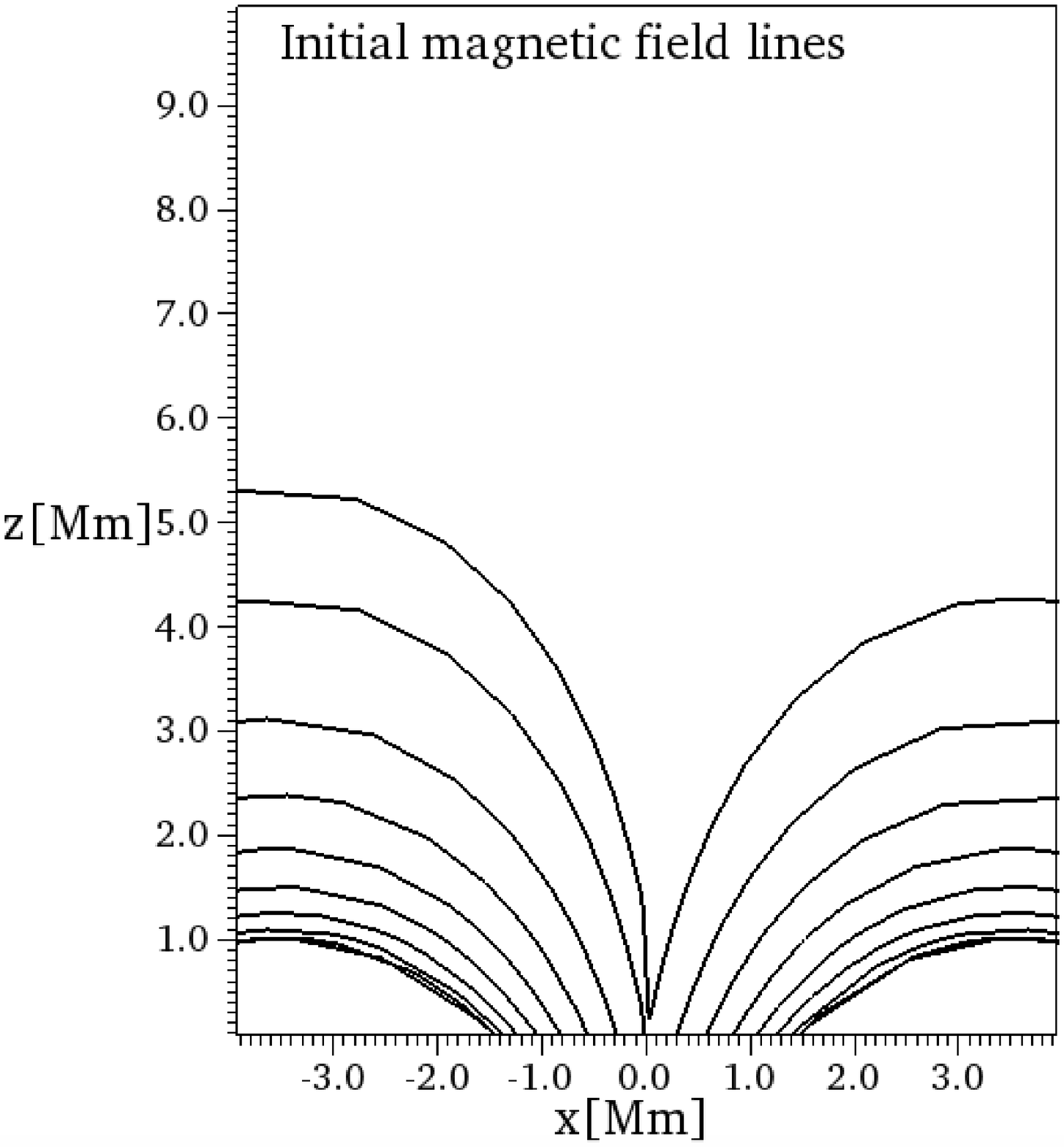}
\caption{\label{fig:caseB1} From left to right we show snapshots of i) logarithm of the temperature in Kelvin, ii) the vertical component of the velocity ($v_z$ km/s) in color with the velocity vector field, iii) the $y$ component of the current density $J_y$(A/m$^2$) at $t=180$ s, iv) on the extreme right we show the magnetic field configuration at initial time. In the Top panel we show the results for Run \#7, where $B_{01}=40,~B_{02}=30$ G. In the Middle we present the results for Run \# 9, where  $B_{01}=40,~B_{02}=20$ G. Finally in the Bottom we show the results for Run \# 13, where  $B_{01}=30,~B_{02}=20$. In all the cases $l_0=3.5$Mm.}
\end{figure*}

In this case we show the results for a more realistic magnetic configuration of the numerical simulations corresponding to the case of non-symmetric magnetic field loops, i.e. when the magnetic strengths of the left and right loops are different ($B_{01} \ne B_{02}$) for the combinations of magnetic field strengths 20, 30 and 40 G and $l_0=~2.5,~3.0,~3.5$ Mm. In order to illustrate the effect of the asymmetry in the formation of jets, we show the results for Runs \# 7, \# 9 and  \# 13 in Fig. \ref{fig:caseB1}. At the top panel we present the result for Run \# 7, that shows the inclination of the jet toward the loop with the weak magnetic field. Similar to the previous case, the top part of the jet exhibits a bulb which appears due to Kelvin-Helmholtz instability \citep{Kuridze_et_al_2016}. This jet reaches a height of 8.5 Mm and a maximum vertical velocity $v_{z,max}\approx 24.65$ km/s at $t=180$ s. Like in the symmetric case, the jet starts to weaken and finally vanishes. At the middle panel of Fig. \ref{fig:caseB1} we show Run \# 9, in this case the jet shows a more significant inclination to the direction of the weaker magnetic field. In this case the jet reaches a height of 8.0 Mm and a maximum velocity $v_z\approx 44.15$ km/s at time $t=180$ s. The inclination is also shown in the $y$ component of the current density $J_y$. At the bottom we show  Run  \# 13, in this case the jet is small due to the weaker magnetic field of the loops. The maximum height of the jet is about 5 Mm and its maximum vertical speed is $v_z\approx 32.14$ km/s at $t=180$s.

The inclined propagation of the jet produces a distortion of the magnetic field lines, that can be seen by comparing the lines at initial time with the lines at the time of the snapshot. The velocity vector field shows the plasma moving in the direction of the weaker magnetic field. The $y$ component of the current density $J_y$ shows the formation of an elongated current sheet, directly related to the elongated shape of the jet and similar shape like those observed by Hinode for instance in Fig. 1 of \citep{Tavabi_et_al_2015b}. We do not show the density of the plasma in these Figures, however its shape is pretty much that of the Temperature profile.

We summarize the results with combinations of the magnetic field configurations presented in Table  \ref{tab:CaseB} for symmetric and non-symmetric magnetic loops. In the same Table we also show the resulting values for the maximum velocity of the plasma along the vertical direction $v_{z,max}$, plasma temperature estimate $T_{inside}$ and density estimate $\rho_{inside}$ measured along the jet.

\begin{table}
\caption{Maximum velocity, temperature and density of jets in terms of the magnetic field strength of the two loops. We include symmetric ($B_{01}=B_{02}$) and non-symmetric configurations ($B_{01} \ne B_{02}$).}
\centering
\resizebox{0.48\textwidth}{!}{
\begin{tabular}{c | c c c c c c}
\hline\hline
Run \#& $B_{01}$ (G) & $B_{02}$ (G) & $l_0$ (Mm) & $v_{z,max}$ (km/s) & $T_{inside}$(K) & $\rho_{inside}$(kg/m$^{3}$)\\ [0.5ex]
\hline
1 & 40  & 40 & 3.5  & 34.1  & 99282 & $1.2\times10^{-10}$  \\
2  & 30  & 30 & 2.5 &  34.8 & 9799 & $4.7\times10^{-10}$  \\
3  & 30  & 30  &  3.5 & 32.5  & 92833 & $2.3\times10^{-10}$ \\
4  & 20  & 20  &  2.5 & 16.9 & 13734  & $1.4\times10^{-10}$  \\
5  & 20  & 20  &  3.0 &  20.5 & 124199  & $2.5\times10^{-10}$   \\
6 & 20  & 20   &  3.5 &  21.3  & 103675 & $3.3\times10^{-11}$  \\
7 & 40  & 30  &  3.5 &  24.7   & 24586 & $2.7\times10^{-11}$  \\ 
8  & 40  & 20  &  3.0  &  78.0 & 88812  & $5.0\times10^{-10}$  \\
9  & 40  & 20  &  3.5  &  44.2  & 60599  & $1.4\times10^{-11}$   \\
11 & 30 & 20  &  2.5  &  58.7 & 71638  & $3.9\times10^{-10}$   \\
12 & 30 & 20  &  3.0  &  49.3 & 96158  & $3.8\times10^{-10}$   \\
13 & 30 & 20  &  3.5  &   32.1 & 49799 & $3.0\times10^{-11}$  \\
[1ex]
 \hline
\end{tabular}}
\label{tab:CaseB}
\end{table}

\begin{figure}
\centering
\includegraphics[width=6.25cm,height=6.25cm]{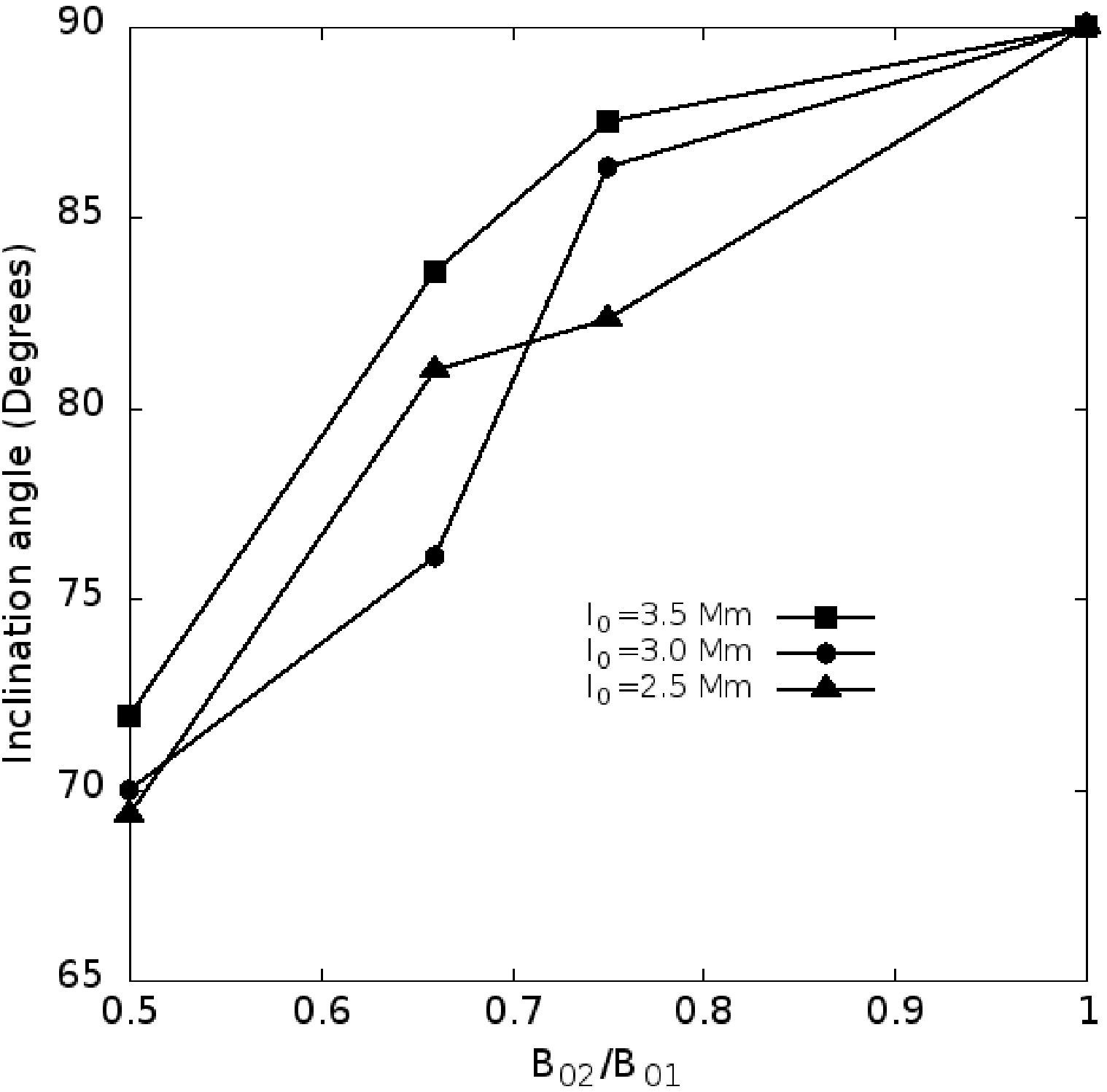}
\caption{\label{fig:Ratio_vs_angle} The inclination angle of the jet as a function of the ratio $B_{02}/B_{01}$ for the cases of two non-symmetric magnetic loops with separations $l_0=3.5,3.0,2.0$ Mm.}
\end{figure}

In the case of two non-symmetric magnetic loops the inclination depends of the magnetic field strength between the two loops as shown in Fig. \ref{fig:caseB1}.  In order to see this dependence more clearly we show the inclination angle of the jet as a function of the ratio $B_{02}/B_{01}$ for different separation parameters in Fig. \ref{fig:Ratio_vs_angle} at the time when the jet reaches the maximum height for each case. 

\begin{figure}
\centering
\includegraphics[width=8.5cm,height=4.0cm]{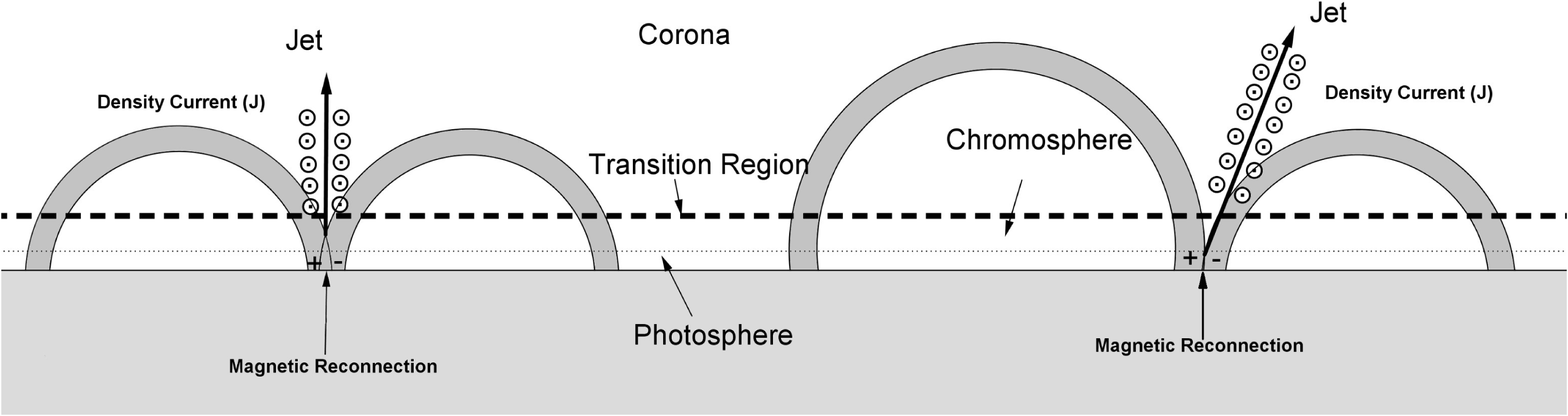}
\caption{\label{fig:process} We show the two configurations of the magnetic coronal loops with foot points at the photospheric level. Initially two magnetic loops close together have opposite polarity which produces magnetic reconnection. In the case of two symmetric magnetic loops (left) the jet appears at the middle of the configuration and moves straight upward until it reaches the solar corona and diffuses later on. In the case of two asymmetric loops (right) the jets appears at the middle of the configuration, but in this case the loop with the stronger magnetic field pushes the jet toward the weak magnetic field and the jet reaches the solar corona. In both cases there is an elongated current sheet represented with the density current $J$ in the perpendicular direction to the jet.}
\end{figure}

\section{Conclusions}
\label{sec:conclusions}

In this paper we present the numerical solution of the equations of the resistive MHD submitted to the solar constant gravitational field, and simulate the formation of narrow jet structures on the interface low-chromosphere and corona. For this we use a magnetic field configuration of two superposed loops in a way that a current sheet is formed that allows the magnetic reconnection process, which in turn accelerates the plasma.

An ingredient of our simulations is that we use a realistic atmospheric model that includes the transition region. The rarefaction of the environment above the transition region helps the acceleration of the plasma. We can summarize our findings in the schematic picture  shown in Fig.\ref{fig:process}.  This rarefied atmosphere allows the formation of a bulb at the top of the jet due to Kelvin-Helmholtz instability \citep{Kuridze_et_al_2016}, which is contained and stabilized by the magnetic field as found in \citep{Flint_2014,Zaqarashvili_2014}. 

We consider symmetric and asymmetric magnetic field configurations. In the symmetric case different jet properties where found in terms of separation and magnetic field strength of the loops. The magnetic field used ranges from 20 to 40 G, leading to the conclusion that the stronger the magnetic field the higher and faster the jet. The separation of the loops' feet was also found to be important, because it determines the thickness of the current sheet that later on produces the magnetic reconnection. With our parameters, a separation with $l_0=4$Mm is so large that no jet is formed anymore. The Temperature within the jet structure is of the order of $10^{4}$ K, which is within the observed range of a cool jet \citep{Nishizuka_et_al_2008}. An illustrative example is that of $B_{01}=B_{02}=40$G and $l_0=3.5$Mm, which shows a height of 7Mm measured from the transition region, and a maximum vertical velocity of $v_z\approx 34$ km s${}^{-1}$, parameters similar to those of Type II spicules  \citep{De_Pontieu_et_al_2007c}. The evolution of these structures indicate that they last about 200 s, which is a lifetime similar to type II spicules. 

In the case of asymmetric magnetic field configurations we also simulated the formation of jets with similar properties of Temperature, velocity and height. These jets show a considerable inclination toward the loop with the weaker magnetic field. We found that the inclination of the jet depends on the magnetic field ratio of the two loops. 

According to the results of this paper, a good model for the formation of realistic jets mimicking type II spicules is to have two magnetic loops close together with opposite polarity. This produces a current sheet at chromospheric level capable to trigger magnetic reconnection. A key ingredient in the process is the inclusion of magnetic resistivity, which is a mechanism consistent with the magnetic reconnection process. Another feature of these jets is that they are based at the level of the transition region, which is characterized by a sharp gradient in density and temperature.

\acknowledgments{Acknowledgments. This research is partly supported by the following grants: Newton Fund -- MAS/CONACyT Mobility Grants Program, Royal Society-Newton Mobility Grant NI160149, CIC-UMSNH 4.9 and CONACyT 258726 (Fondo Sectorial de Investigaci\'on para la Educaci\'on). V. F. would like to acknowledge STFC for support  received. The simulations were carried out in the IFM Draco cluster funded by  CONACyT 106466 and in the Sciesmex cluster at IG-UNAM.


\bibliographystyle{yahapj}

\end{document}